\documentclass[a4paper,11pt]{spieman}  % use this instead for A4 paper

\usepackage{amsmath,amsfonts,amssymb}
\usepackage{graphicx}
\usepackage{setspace}
\usepackage{tocloft}
\usepackage{lineno}
\usepackage{enumitem}
\usepackage{parskip} % for the noitemsep command
\usepackage{algorithm}
\usepackage{algorithmicx}
\usepackage{algpseudocode}
\title{SpaceWire-based Data Acquisition Network for the Solar Flare Sounding Rocket Experiment FOXSI-4 and FOXSI-5}

\author[a,b,c*]{Shunsaku Nagasawa}
\author[d]{Athanasios Pantazides}
\author[d]{Kristopher Cooper}
\author[e]{Riko Shimizu}
\author[a]{Savannah Perez-Piel}
\author[b,c]{Takahiro Minami}
\author[d]{Yixian Zhang}
\author[a]{Hunter Kanniainen}
\author[f]{Shin Watanabe}
\author[c,g]{Tadayuki Takahashi}
\author[h,i]{Noriyuki Narukage}
\author[a]{Juan Camilo Buitrago Casas}
\author[d]{Lindsay Glesener}
\affil[a]{Space Sciences Laboratory, University of California, Berkeley, 7 Gauss Way, Berkeley, CA 94720, USA}
\affil[b]{Department of Physics, The University of Tokyo, 7-3-1 Hongo, Bunkyo, Tokyo 113-0033, Japan}
\affil[c]{Kavli Institute for the Physics and Mathematics of the Universe (Kavli IPMU, WPI), The University of Tokyo, 5-1-5 Kashiwanoha, Kashiwa, Chiba 277-8583, Japan}
\affil[d]{University of Minnesota, Twin Cities, Minneapolis, Minnesota, USA}
\affil[e]{Department of Space and Astronautical Science, The Graduate University for Advanced Studies, SOKENDAI, 3-1-1, Yoshinodai, Chuo-ku Sagamihara, Kanagawa 252-5210, Japan}
\affil[f]{Institute of Space and Astronautical Science, Japan Aerospace Exploration Agency (ISAS/JAXA), 3-1-1 Yoshinodai, Chuo-ku, Sagamihara, Kanagawa 252-5210, Japan}
\affil[g]{International Center for Quantum-field Measurement Systems for Studies of the Universe and Particles (QUP, WPI), KEK, Ibaraki 305-0801, Japan}
\affil[h]{National Astronomical Observatory of Japan, 2-21-1 Osawa, Mitaka, Tokyo 181-8588, Japan}
\affil[i]{Department of Astronomical Science, The Graduate University for Advanced Studies, SOKENDAI, 2-21-1 Osawa, Mitaka, Tokyo 181-8588, Japan}

\setlist{noitemsep} % globally set lists to only have single linebreak between items (itemize and enumerate)

\cftpagenumbersoff{figure}
\cftpagenumbersoff{table} 
\begin{document} 
\maketitle

\begin{abstract}
We developed a SpaceWire-based data acquisition (DAQ) system for the FOXSI-4 and FOXSI-5 sounding rocket experiments, which aim to observe solar flares with high sensitivity and dynamic range using direct X-ray focusing optics.
The FOXSI-4 mission, launched on April 17, 2024, achieved the first direct focusing observation of a GOES M1.6 class solar flare with imaging spectroscopy capabilities in the soft and hard X-ray energy ranges, using a suite of advanced detectors, including two CMOS sensors, four CdTe double-sided strip detectors (CdTe-DSDs), and a Quad-Timepix3 detector.
To accommodate the high photon flux from a solar flare and these diverse detector types, a modular DAQ network architecture was implemented based on SpaceWire and the Remote Memory Access Protocol (RMAP). 
This modular architecture enabled fast, reliable, and scalable communication among various onboard components, including detectors, readout boards, onboard computers, and telemetry systems.
In addition, by standardizing the communication interface and modularizing each detector unit and its associated electronics, the architecture also supported distributed development among collaborating institutions, simplifying integration and reducing overall complexity.
To realize this architecture, we developed FPGA-based readout boards (SPMU-001 and SPMU-002) that support SpaceWire communication for high-speed data transfer and flexible instrument control.
In addition, a real-time ground support system was developed to handle telemetry and command operations during flight, enabling live monitoring and adaptive configuration of onboard instruments in response to the properties of the observed solar flare.
The same architecture is being adopted for the upcoming FOXSI-5 mission, scheduled for launch in 2026.
\end{abstract}

% Include a list of up to six keywords after the abstract

% Include email contact information for corresponding author
{\noindent \footnotesize\textbf{*}Corresponding Author: Shunsaku Nagasawa,  \linkable{nshunsaku@berkeley.edu} }

% \begin{spacing}{2}   % Remove for edit: use double spacing for rest of manuscript
%

\section{Introduction}
\label{sect:intro}  
The Focusing Optics X-ray Solar Imager (FOXSI)\cite{krucker2013focusing} is the first solar-dedicated sounding rocket mission to perform direct focusing imaging observations of the Sun using high-resolution Wolter-I X-ray focusing optics combined with fine-pitch focal plane detectors.
By employing the direct imaging method, FOXSI can achieve a greater sensitivity and imaging dynamic range\cite{piana2022hard, krucker2014first} than those of previous missions based on indirect imaging methods (e.g., the Reuven Ramaty High-energy Solar Spectroscopic Imager satellite\cite{lin2003reuven, hurford2003rhessi}). This significant advancement enables the detection of much fainter solar X-ray sources.
With these capabilities, FOXSI facilitates detailed investigations of small-scale solar phenomena, such as active regions, microflares, and nanoflares, which are essential to understanding the long-standing problem of coronal heating\cite{judge2024problem}. FOXSI’s first three successful flights (FOXSI-1, -2, and -3, launched in 2012, 2014, and 2018, respectively) targeted relatively quiet regions of the Sun, demonstrating its enhanced sensitivity and leading to a number of important scientific discoveries\cite{krucker2014first, ishikawa2017detection, athiray2020foxsi, vievering2021foxsi, buitrago2022faintest}.
Building on these successes, the focus of subsequent FOXSI observations was placed on medium- and large-scale solar flares. These observations aim to deepen our understanding of flare-related particle acceleration by clarifying where particles are energized in the corona, how energetic electrons propagate and lose energy, and revealing the mechanisms by which accelerated particles escape into interplanetary space.

The fourth launch, FOXSI-4, was proposed and selected under NASA's Heliophysics Low Cost Access to Space (H-LCAS) program\cite{buitrago2021foxsi}.
FOXSI-4 was designed to demonstrate the capabilities of the first-ever direct focusing observation of mid-to-large class solar flares ($\geq$ GOES C5 class) in soft X-rays and hard X-rays.
FOXSI-4 was successfully launched from the Poker Flat Research Range in Fairbanks, Alaska, on April 17, 2024, and observed a mid-class (GOES M1.6 class) solar flare. Following this success, FOXSI-5, a re-flight mission based on the FOXSI-4 configuration, was selected and is scheduled for launch in Spring of 2026 from the White Sands Missile Range (WSMR) in New Mexico.

In this paper, we present the design, implementation, and performance evaluation of the data acquisition system developed for FOXSI-4, along with its extension for the upcoming FOXSI-5 mission.
The system is designed to meet the demands of high count rate solar flare observations and the integration of multiple detector types onboard a sounding rocket platform.
There are two primary requirements that motivated the development of a new data acquisition architecture for FOXSI-4.

First, FOXSI-4 flew as part of a solar flare campaign,\cite{savage2022first} with a launch triggered by the onset of a mid-to-large class of solar flare.
Consequently, the expected photon flux is significantly higher than in previous FOXSI missions, which focused on quiet solar regions. 
% For example, the unattenuated count rate from an M-class flare can easily exceed one million photons per second.
To accommodate this high photon rate, the data acquisition system must support high-rate readout and be capable of efficiently processing the data volume.
Moreover, since the incident photon flux can vary by orders of magnitude depending on the flare class, the system must be able to adjust its operating parameters accordingly.
Second, FOXSI-4 features seven telescopes with more diverse optics modules and detectors than previous flights, including four hard X-ray CdTe double-sided strip detector (CdTe-DSD)\cite{nagasawa2023wide, nagasawa2025imaging}, two soft X-ray CMOS sensors\cite{shimizu2024evaluation}, and one hard X-ray Quad-Timepix3 CdTe detector\cite{buitrago2022quad}.
Since previous FOXSI launches mainly employed the same type detectors of Si-DSDs and CdTe-DSDs, which were developed and provided by ISAS/JAXA and Kavli IPMU/The University of Tokyo, we used the same control and readout system based on a simple FPGA board. However, the introduction of new types of detectors requires the system to simultaneously support multiple types of instruments with different operational and data handling requirements.

To meet these requirements, we developed a modular onboard data processing system based on the standard communication interface ``SpaceWire" and the ``Remote Memory Access Protocol (RMAP)".
As part of this system, we developed an FPGA-based readout board, SPMU-001 and SPMU-002, to support high-speed data acquisition and flexible instrument control.
In Section\,\ref{sec:daq_foxsi4}, we present an overview of the data acquisition network for FOXSI-4, including the design of the SPMU-001 FPGA board. Sections\,\ref{sec:formatter} through \ref{sec:timepix_daq} describe the individual components of the data acquisition system in detail. Finally, Section\,\ref{sec:gse} introduces the ground control system used for system configuration and data monitoring.

\section{Overview of Data Acquisition Network} \label{sec:daq_foxsi4}
For FOXSI-4 and FOXSI-5, fast, reliable, and modular communication between various onboard instruments, including multiple types of detectors, readout boards, onboard computers and telemetry systems, is essential.
To support this, the need for a standardized and scalable data communication system is critical.
In addition, by modularizing each detector unit and its associated electronics, the system allowed for distributed development among collaborating institutions, thereby streamlining integration and reducing overall complexity.

To address these requirements, FOXSI-4 adopted SpaceWire, a bi-directional, high-speed serial communication protocol specifically designed for spacecraft. SpaceWire has been standardized mainly by ESA, NASA, JAXA, and Roscosmos\cite{ecssSpaceWire} and is widely used in various space missions\cite{ozaki2010spacewire, hihara2020onboard, rakow2003reliable}. The use of a standard protocol makes it possible to shorten the development time and cost, by using a consistent interface for each onboard instrument. 
In combination with the Remote Memory Access Protocol (RMAP)\cite{ecssRMAP}, which allows direct read/write access to remote memory over the SpaceWire network, the system achieves high flexibility and low-latency data acquisition. 

\begin{figure}[h]
  \begin{center}
  \includegraphics[width=1.0\hsize]{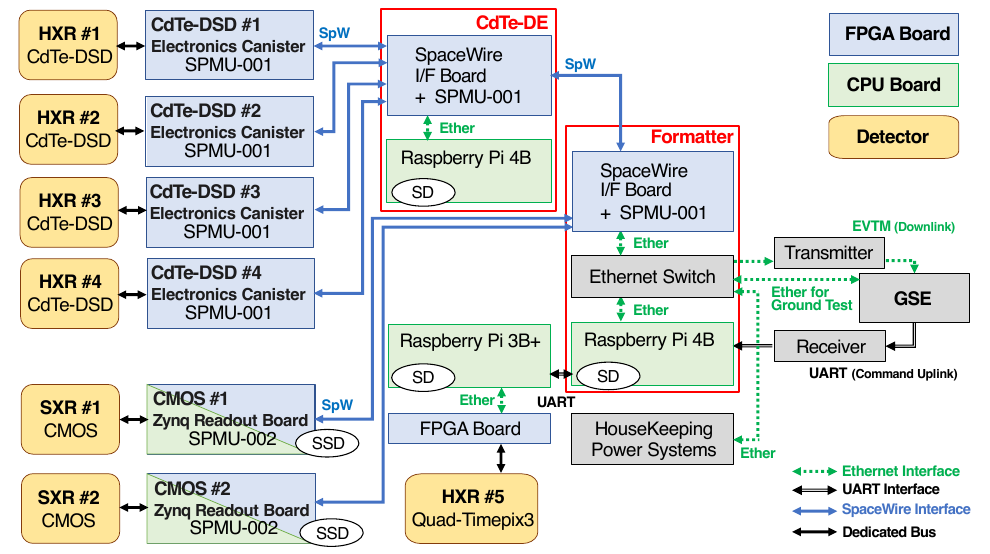}
  \caption{Block diagram of the SpaceWire data acquisition network and data flow in FOXSI-4 and FOXSI-5. Yellow, blue, and green boxes represent the detectors, readout FPGA boards, and CPU boards (Raspberry Pi 4B or 3B+), respectively. Arrows indicate the flow of data and commands: green dashed arrows represent the Ethernet interface, black double-headed arrows represent the UART interface, and blue arrows represent the SpaceWire interface.}
  \label{data_network}
  \end{center}
\end{figure}

Fig\,\ref{data_network} shows a network diagram of the data flow in FOXSI-4.
The Formatter (Section \ref{sec:formatter}) is responsible for telemetry communication with the ground system. It collects QuickLook (QL) data from each detector module, including the CdTe-DSD, CMOS, and Timepix detectors, and transmits them to the ground.
Communication with each module is conducted in \textit{pull} mode, in which data are retrieved based on requests from the Formatter.
In addition, the Formatter sends control commands to each module, enabling remote configuration and operation during flight.
The data downlink is transmitted to the ground through an EVTM (Ethernet Via Telemetry) system, and the maximum rate is limited to $\sim$ 20 Mbps. The downlinked data is then processed by the Ground Support Equipment (GSE, Section \ref{sec:gse}) computer, and QL lightcurve, spectra, and housekeeping information are monitored.

Commands are uplinked through UART serial communication on a 1.2 kbaud line (see Ref.~\citenum{burth2023nasa} for details) from the GSE. Commands can be used to start or end of observations, change  the bias voltage, and configure other parameters.
The Formatter and GSE software has been developed by University of Minnesota.

The four CdTe-DSDs are read out by the CdTe-DSD Data Electronics (CdTe-DE, Section\,\ref{sec:CdTe-DE_overview}), which is responsible for data acquisition as well as responding to and acting on control commands from the Formatter.
Each detector is readout by an ``Electronics Canister", which includes a new FPGA board, the SPMU-001 (Section\,\ref{sec:spmu001}). 
Observation data from each CdTe-DSD are first written to the SDRAM on the SPMU-001 in each canister and then collected by the CdTe-DE.
All observation data are subsequently stored on the SD card of the Raspberry Pi 4 of the CdTe-DE, which is physically retrieved after launch and payload recovery.
In addition, a part of the data is written to SDRAM on the SPMU-001 in CdTe-DE for QL and backup, then transmitted to the ground system via telemetry.
The CdTe-DE hardware and software were jointly developed by Kavli IPMU, and ISAS/JAXA.
 
The data acquisition and control of the two CMOS sensors, as well as the generation of QL data, are handled by a dedicated data acquisition board, SPMU-002 (Section\,\ref{sec:cmos_daq}).
SPMU-002 is an upgraded version of the ZDAQ board in FOXSI-3 \cite{ISHIKAWA2018191}, with a new Zynq UltraScale+ MPSoC (MultiProcessor System on a Chip). 
It also includes a SpaceWire interface that allows external devices to access its onboard SDRAM.
With SPMU-002, all observation data are stored on an onboard SSD, while a subset is written to SDRAM for telemetry use.
The CMOS DAQ module has been developed by NAOJ.

For FOXSI-4, the Timepix3 data electronics system was developed by Space Sciences Laboratory (SSL) at UC Berkeley (Section \ref{sec:timepix_daq}).
It consists of an AMD Kintex-7 FPGA and a Raspberry Pi 3B+, connected via Ethernet with support for 9000-byte jumbo frames.
The FPGA handles event packetization, while the Raspberry Pi manages telemetry and data storage.
All science data packets are written to local storage in PCAP format, and a subset of QL telemetry, including housekeeping and  basic event statistics such as average energy and count rate, is transmitted to the ground during flight.

The Timepix3 system does not use SpaceWire unlike other subsystems in the FOXSI-4 payload, but instead relies on a UART interface for telemetry and command exchange with the Formatter.
The Timepix3 system already included a developed readout system at the outset of the FOXSI-4 project. However, the system did not yet have flight heritage, so the design of the Timepix-Formatter interface prioritized use of existing readout systems and protocols, and minimized the telemetry volume allocated to Timepix3, in order to reduce technical and schedule risk. These constraints motivated the decision to implement serial-based communication rather than utilize the existing SpaceWire system. A simple duplex UART interface for Timepix–Formatter telemetry and commanding was implemented with minimal modification of the Timepix FPGA core, in contrast to a relatively resource-intensive SpaceWire RMAP node implementation.

\subsection{Timing and Clock Synchronization}

Clock synchronization in the FOXSI-4 readout system was designed around a 1 Hz pulse-per-second (PPS) signal received from the rocket GPS system. The rising edge of the PPS signal is aligned to each UTC second by the GPS receiver. This signal was distributed to each onboard readout system in parallel, including the Formatter, CdTe-DE, each CMOS system, and the Timepix FPGA, with the intention of reading the rising edge time and recording a local clock counter in a log within each system. This approach would allow each readout system to align its local clock counter to an absolute timescale, and enable post-flight alignment of all detector data. 

However, with limited time to validate the results of this PPS architecture during the payload integration, only the CdTe-DE flew with its PPS hardware and software intact. As a result, the DE system successfully recorded 1 Hz PPS timestamps in a flight log file, including their alignment with the local 64 Hz SpaceWire timecode. But without other onboard systems utilizing the PPS, it is of little value for clock alignment.  For FOXSI-5, the PPS architecture has been validated during integration tests, and the team expects to benefit from local and absolute clock synchronization from this system in flight. 

\subsection{SPMU-001 FPGA Board} \label{sec:spmu001}
The SPMU-001 is a compact, general-purpose FPGA board designed for physical measurement applications. It was jointly developed by Shimafuji Electric Incorporated and Kavli IPMU.
The board integrates an AMD Spartan-7 FPGA and 128 MB of DDR2-SDRAM, along with high-speed communication interfaces such as SpaceWire and Gigabit-Ethernet within a Raspberry Pi-sized board.
To facilitate system integration, the SPMU-001 provides a 40-pin header compatible with the Raspberry Pi 4, allowing direct power supply from the SPMU-001 to the Pi.
Furthermore, since SpaceWire is not natively supported by the Raspberry Pi, the SPMU-001 provides a SpaceWire–Ethernet bridge interface, allowing software running on the Raspberry Pi to communicate with SpaceWire devices by sending and receiving RMAP packets over a standard TCP connection. 

An AXI–SpaceWire RMAP bridge is also implemented to support direct access to onboard SDRAM via RMAP, enabling data to be acquired over the network without requiring CPU involvement.
For expanded SpaceWire connectivity, an additional interface board, SPMU-001-SpW, was developed for FOXSI-4. This daughterboard mounts on top of the SPMU-001 and provides the necessary LVDS drivers to support up to six physical SpaceWire ports.
Table~\ref{tab:spw_router_list} summarizes the assignment and function of each port in the ten-port SpaceWire router configuration realized by the combination of the SPMU-001 and the SPMU-001-SpW.

\begin{table}[htbp]
  \centering
  \caption{SpaceWire Router Port configurations of SPMU-001 with SPMU-001-SpW}
  \label{tab:spw_router_list}
  \begin{tabular}{l|l}
    SpW Router Port Number & Function \\ \hline
     0            & Configuration Register Port \\
     1            & SpaceWire I/F, SpW Physical Port 1 \\
     2            & AXI–RMAP Bridge \\
     3, 4, 5          & SpaceWire–Ethernet Bridge (TCP/IP, 3 Ports) \\
     6, 7, 8, 9, 10   & SpaceWire I/F, SpW Physical Ports 2, 3, 4, 5, 6 \\
  \end{tabular}
\end{table}

\begin{figure}[H]
  \begin{center}
  \includegraphics[width=1.0\hsize]{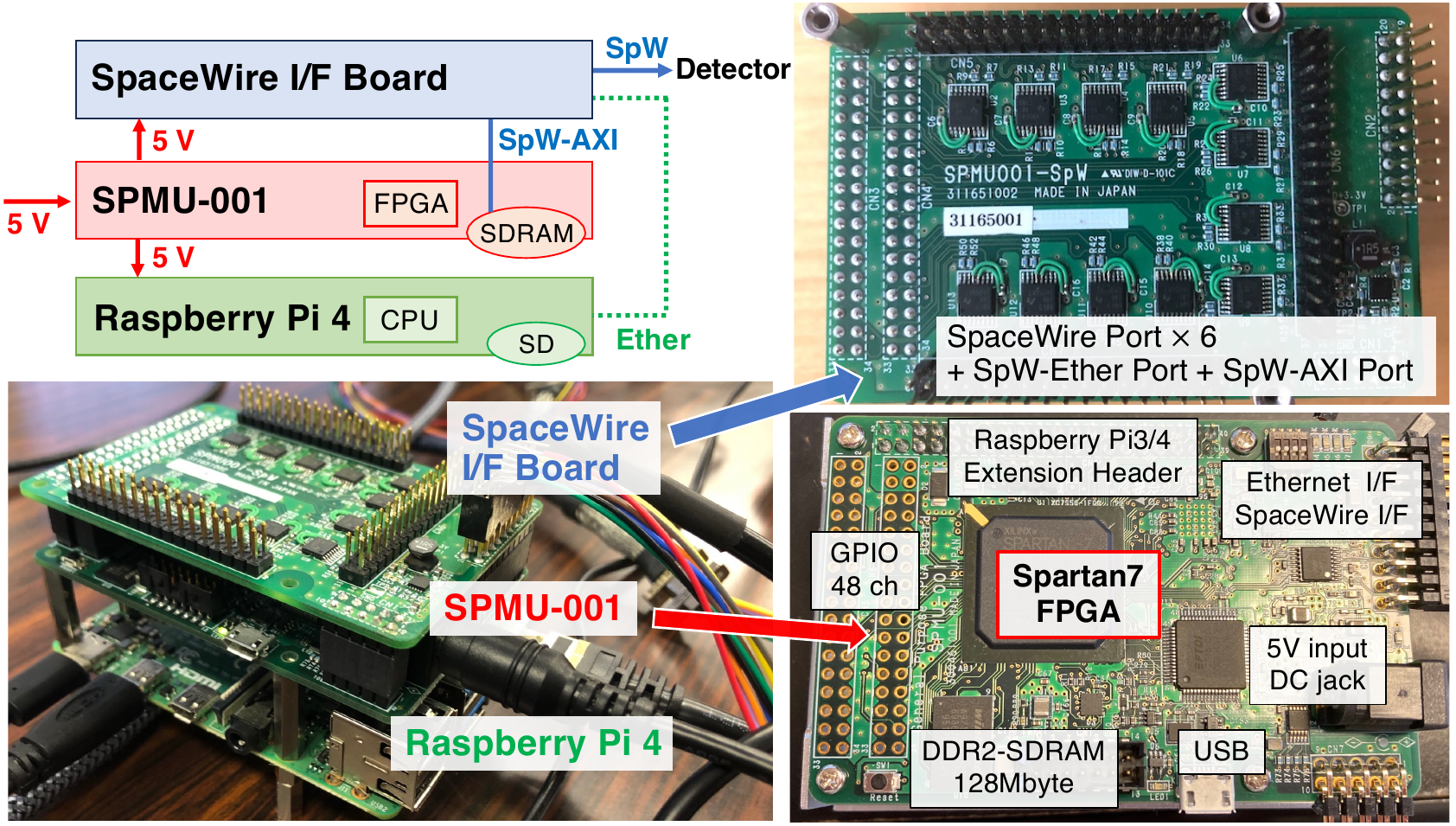}
  \caption{Pictures of the SpaceWire extension I/F board (SPMU-001-SpW), the SPMU-001 FPGA board, and the Raspberry Pi 4. All three boards can be stacked together, and the power can also be supplied directly from the SPMU-001 to Raspberry Pi 4 and SPMU-001-SpW.}
  \label{data_daq}
  \end{center}
\end{figure}

For FOXSI-4 and FOXSI-5, the SPMU-001 is used as the DAQ board responsible for controlling each CdTe-DSD.
The detector module is placed in a 10 cm-diameter circular housing, while the associated data and power circuit board is integrated into the cylindrical unit referred to as the Electronics Canister (see Fig\,\ref{data_network} and Fig\,2 in Ref.~\citenum{nagasawa2025imaging}). 
% The power board inside the canister includes a DC–DC converter that steps down the 28 V input to 5 V to supply power to the DAQ board. 
% In addition, it generates and distributes the required voltages for ASICs and supplies the bias voltage necessary for detector operation.
The DAQ board controls the detectors and stores the acquired data in onboard SDRAM.
The configuration of ASIC parameters and the selection of measurement modes can be performed by modifying registers in SDRAM via the SpaceWire/RMAP protocol.
Additionally, the three-board set consisting of the SPMU001-SpW, SPMU-001, and Raspberry Pi 4 (as shown in Fig\,\ref{data_daq}) is utilized in two locations within the architecture: once as the CdTe-DE (see Section \ref{sec:CdTe-DE_overview}), and once as the Formatter for telemetry communication with the ground system  (see Section \ref{sec:formatter}).

\section{Formatter} \label{sec:formatter}
During flight, data from the seven FOXSI detectors are digitized for onboard mass storage. A fraction of this data is aggregated centrally and transmitted to the ground for live monitoring via a device known as the \textit{Formatter}. The Formatter supports a variety of physical interfaces that enable communication with the various detector readout systems, housekeeping and power management systems, telemetry system, and command uplink interface. These interfaces are outlined in Fig\,\ref{data_network}.

    % \begin{figure}[h!]
    %     \centering
    %     \includegraphics[width=\textwidth]{experiment_data_legend.pdf} 
    %     \caption{Formatter data interfaces}
    %     \label{fig:formatter_ports}
    % \end{figure}
	
Table \ref{tab:interfaces} identifies the Open Systems Interconnection (OSI) model stack for each Formatter interface. The variety of interface requirements drives the hardware design of the system.
    \begin{table}[h!]
        \centering
            \caption{OSI components of Formatter interfaces.}
        \begin{tabular}{c|cccc}
             & Physical layer & Data link layer & Network layer & Transport layer \\ 
            \hline 
            CdTe-DSDs & LVDS & SpaceWire & RMAP & RMAP \\ 
            CMOS sensor & LVDS & SpaceWire & RMAP & RMAP \\ 
            Timepix & RS-232 & 8N1 UART & --- & --- \\ 
            Power/Housekeeping & 10BASE-T & Ethernet & IPv4 & TCP \\ 		
            Downlink & 10BASE-T & Ethernet & IPv4 & UDP \\ 
            Uplink & RS-232 & 8N1 UART & --- & --- \\ 
        \end{tabular} 
        \label{tab:interfaces}
    \end{table}
    % \begin{table}[h!]
    %     \centering
    %         \caption{OSI components of Formatter interfaces.}
    %     \begin{tabular}{ccccc}
    %         \textbf{Endpoint system} & \textbf{Physical layer} & \textbf{Data link layer} & \textbf{Network layer} & \textbf{Transport layer} \\ 
    %         \hline 
    %         \textbf{CdTe-DSD} & LVDS & SpaceWire & RMAP & --- \\ 
    %         \textbf{CMOS} & LVDS & SpaceWire & RMAP & --- \\ 
    %         \textbf{Timepix} & RS-232 & 8N1 UART & --- & --- \\ 
    %         \textbf{Power/housekeeping} & 10BASE-T & Ethernet & IPv4 & TCP \\ 		
    %         \textbf{Downlink} & 10BASE-T & Ethernet & IPv4 & UDP \\ 
    %         \textbf{Uplink} & RS-232 & 8N1 sUART & --- & --- \\ 
    %     \end{tabular} 
    %     \label{tab:interfaces}
    % \end{table}

\subsection{Hardware Design} \label{sec:formatter-hw}

The Formatter is composed of a main processor, an Ethernet switch, and an Ethernet-to-SpaceWire converter. The processor is the Raspberry Pi 4B, loaded with a 32 GB SD card, and running Raspbian (a variant of the Debian Linux). This single-board computer was chosen early in the Formatter development due to its popularity, heritage in spaceflight projects, and availability of documentation and community resources for development. 
The Raspberry Pi includes one physical Ethernet port, but is required to communicate with multiple systems via Ethernet, so a custom 5-port Ethernet switch was developed. This board was also designed to the Raspberry Pi form factor and 40-pin header, and is based on the Microchip KSZ8567S managed Ethernet switch. This switch accommodates connections between the Raspberry Pi, SPMU-001, power/housekeeping system, and downlink system. 
% SpaceWire is not natively supported by the Raspberry Pi. To facilitate communication with SpaceWire devices, an SPMU-001 was procured from Shimafuji Electric Corporation. This device exposes the SpaceWire RMAP over a network interface \cite{ecssRMAP, ecssSpaceWire}. The SPMU-001 is designed to a Raspberry Pi form factor and 40-pin header, so integration of the two boards is convenient. When connected, the Raspberry Pi can treat the SPMU-001 as a TCP connection, and application-layer software on the Pi can be used to receive and transmit SpaceWire RMAP packets which are encapsulated in TCP packets on the network. An additional interface card called the SPMU-001-SPW sits on top of the SPMU-001 and carries the necessary LVDS drivers to support five physical SpaceWire interfaces.

\begin{figure}[h!]
        \centering
        \includegraphics[width=1\textwidth]{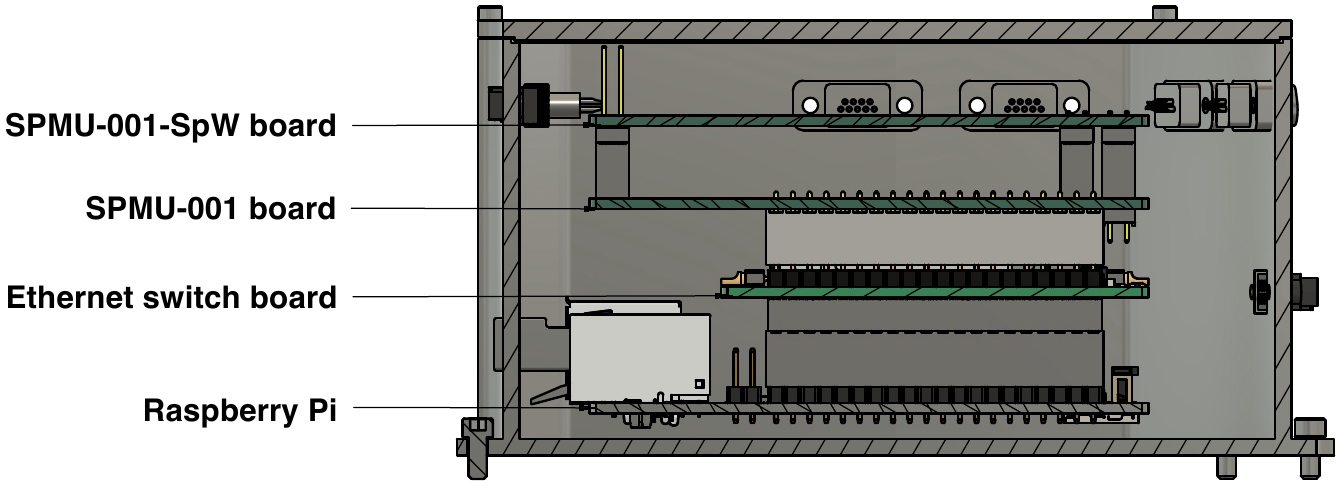} 
        \caption{Formatter board stack.}
        \label{fig:formatter_stack}
    \end{figure}

Fig \ref{fig:formatter_stack} illustrates the mechanical board stackup. This board stack includes the Raspberry Pi/SPMU-001/SPMU-001-SpW triad described in Section \ref{sec:spmu001} and Fig \ref{data_daq} for SpaceWire support. Pigtails link panel-mount connectors to the boards to provide secure mechanical connections to external systems. This configuration survived unit-level and full-system vibration testing campaigns.
Following a failed thermal/vacuum test, custom low-cost heat straps were built for the Formatter stack to sink heat from the Raspberry Pi processor and SPMU-001 FPGA into the mechanical box. These were built by crimping compression lugs onto short pieces of grounding braid. One lug was bonded to the hot processor using a thermally conductive epoxy, and the other terminal was fastened to the mechanical box wall. These straps helped maintain the core temperature of the Raspberry Pi below 60$^\circ$C while operating in vacuum.

\subsection{Software Design} \label{sec:formatter-sw}
The Formatter software\cite{Formatter_soft} polls each detector subsystem in a round-robin loop. After receiving data from each system, data is sent to the telemetry system for downlink. Prior to polling each system, the software checks the uplink device for new uplink commands. These commands are sorted into a FIFO buffer for each detector subsystem. When the loop arrives at a system with an uplink command available in the buffer, the Formatter translates the uplink command code into a format that the system expects, then sends it.

The Formatter software is written in C++ and relies on the Asynchronous Input/Output (ASIO) library packaged by Boost\cite{Boost_Asio} for non-blocking I/O support. This library exposes a unified, abstract API for both network sockets and serial ports, simplifying development and test design. It enables non-blocking read/write access to communication ports, which increases data throughput and provides a mechanism for ports to timeout if communication errors occur.

\subsubsection{Downlink Data Handling} \label{sec:formatter-sw-downlink}
During flight, a subset of all detector data is telemetered to the ground for live monitoring. Each detector subsystem defines different telemetry data products, with unique sizes, structures, and access methods. Table \ref{tab:downlink-products} lists available telemetry data products for each detector system.
To handle the heterogeneous telemetry data products from detector systems, the Formatter software implements layers of a network stack that allows the details of each subsystem's data to be abstracted away.
    % \begin{table}[h!]
    %     \centering
    %         \caption{Telemetry data products for each onboard detector system.}
    %     \begin{tabular}{cccc}
    %         \textbf{Detector system} & \textbf{Housekeeping} & \textbf{Photon-counting data} & \textbf{Quick-look images} \\
    %         \hline 
    %         \textbf{CdTe DE} & $\bullet$ &  &  \\ 
    %         \textbf{CdTe detectors} & $\bullet$ & $\bullet$ &  \\ 
    %         \textbf{CMOS sensors} & $\bullet$ & $\bullet$ & $\bullet$ \\ 
    %         \textbf{Timepix} & $\bullet$ &  &  \\ 
    %     \end{tabular} 
    %     \label{tab:downlink-products}
    % \end{table}
    \begin{table}[h!]
        \centering
            \caption{Telemetry data products for each onboard detector system.}
        \begin{tabular}{c|ccc}
             & Housekeeping & Photon-counting data & Quick-look images \\
            \hline 
            CdTe-DE & $\bullet$ &  &  \\ 
            CdTe-DSDs & $\bullet$ & $\bullet$ &  \\ 
            CMOS sensors & $\bullet$ & $\bullet$ & $\bullet$ \\ 
            Timepix & $\bullet$ &  &  \\ 
        \end{tabular} 
        \label{tab:downlink-products}
    \end{table}

\begin{figure}[h!]
        \centering
        \includegraphics[width=0.8\textwidth]{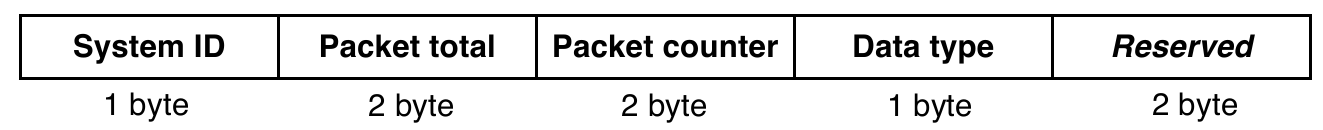} 
        \caption{Header for all telemetry packets.}
        \label{fig:downlink-header}
    \end{figure}

The available photon event lists and QL data products are larger than the network maximum transmissible unit (MTU), so the Formatter implements a custom application-layer fragmentation scheme. In this scheme, telemetry products larger than the network MTU (1500 bytes) are fragmented into packets, each less than the MTU in size, and each packet is prepended with an eight-byte header (see Fig \ref{fig:downlink-header}). 
In the header, the first byte (``system ID'') encodes the onboard system that produced the telemetry data and the fifth byte (``data type'') encodes the type of telemetry data (e.g. housekeeping vs. photon list vs. QL data). The ``packet total'' field counts the number of packets needed to complete the original frame of detector data, and ``packet counter'' is an index for the specific packet into that frame. This enables reconstruction of complete detector frames on the ground over an MTU-limited downlink. 

The radio downlink interface is abstracted on both the sender (rocket-side) and receiver (GSE-side) ends as a UDP interface, which is a connectionless, unreliable, unordered transport-layer protocol.
The Formatter's application-layer fragmentation scheme has an additional benefit of adding packet order information to the protocol. Although UDP does not guarantee that packets will be delivered in the same order they were sent, the eight-byte header does enable ordered reassembly of complete frames on the ground.

During the FOXSI-4 flight and pre-flight testing, the Formatter achieved a bulk downlink rate of $\sim$1.25 Mbps. For FOXSI-5, small software changes have increased the bulk downlink rate to $\sim$4 Mbps. This is achieved through two adjustments: tuning of onboard timeout delays, i.e. less idle waiting for new detector data, if none is yet available; and differential setting of the onboard vs. telemetry MTU value to minimize downlink frame fragmentation while maximizing packet size. 

\subsubsection{Uplink Command Handling} \label{sec:formatter-sw-uplink}
% Uplink commands were designed with the following goals in mind:
%     \begin{enumerate}
%         \item fixed-width commands
%         \item subsystem-addressable commanding
%     \end{enumerate}
Uplink commands were designed with two goals in mind: (1) to use fixed-width commands, and (2) to support subsystem-addressable commanding.
Goal (1) is due to ease of implementation and the short duration of a sounding rocket flight, where there is no time to validate and uplink command \textit{scripts}, as is common for long-duration orbital missions. Goal (2) is to allow detector subsystems to tailor their uplink commands to their operational needs. The Formatter software addresses these by using a two byte-wide uplink command field. The first byte addresses the target system for the command, and the second byte identifies the system-specific command. A lookup table of systems and commands for each system is queried in the Formatter to translate the raw two byte uplink command into a binary format that is intelligible to the target detector system.

Upon receiving an uplink command (over a UART interface to a radio), the Formatter validates the command, then routes it to a queue for the specific onboard system the command is targeting by inspecting the first byte of the command. When the main loop of the Formatter software reaches a given system, its uplink command queue is popped and new uplink commands are delivered to the system, with any replies forwarded to the ground via the telemetry link.

\subsubsection{Configuration Management} \label{sec:formatter-sw-config}
Configuration of the Formatter hardware and interface definitions changed frequently during pre-integration testing. To avoid recompiling the Formatter software at each interface definition change, a global configuration file is used to populate key interface parameters when the software is launched. This allowed different IP addresses, uplink commands, housekeeping data structures, baud rates, etc. to be tested quickly with only configuration file changes, and helped separate application logic from interface definition.

This configuration file was formatted as a JSON file, and referenced separate command definition files for each subsystem. The configuration data was version controlled in a separate Git repository from the main Formatter code. The same configuration JSON file was also ingested by the GSE software to guarantee compatibility between the Formatter and GSE softwares. The Formatter and GSE share a large interface surface, comprising
	\begin{itemize}
		\item All uplink commands (commands for each system, display names, serialization of commands for uplink, and command translation information for Formatter);
		\item Uplink UART interface definition;
		\item Downlink UDP Ethernet interface definition;
		\item Downlink packet and data frame definitions (to reconstruct coherent frames of detector data from individual packets).
	\end{itemize}
The use of a common configuration file reduced the effort needed to keep this large interface functioning correctly while making changes during the AI\&T phase of the project.

\section{CdTe-DSD Data Electronics (CdTe-DE)}\label{sec:CdTe-DE_overview} 
To support high-throughput and large-volume data acquisition during solar flare observations, and to enable flexible control of observation settings depending on flare intensity, we developed a new CdTe-DE for FOXSI-4 and FOXSI-5. 
The primary functions of CdTe-DE are as follows:
\begin{itemize}  \setlength{\parskip}{0cm} \setlength{\itemsep}{0cm} 
  \item Retrieve data from four CdTe-DSDs simultaneously and save to an SD card on Raspberry Pi 4;
  \item Extract a subset of the data and store it in the SDRAM, which is extracted and sent to the GSE by the Formatter as QL data;
  \item Control the CdTe-DSDs in response to SpaceWire command packets received from Formatter.
\end{itemize}
To achieve this functionality, we developed a new FPGA board SPMU-001 (see Section \ref{sec:spmu001}) and C++-based software running on Raspberry Pi 4 (see Section \ref{sec:struct_de}).
This system enables flexible detector control, efficient data acquisition, and real-time communication with the Formatter via SpaceWire. The overall design was optimized to support high-rate observations expected in solar flares, including robust handling of readout mode, low-deadtime operation, and buffered data management. Performance evaluations of the system, including throughput optimization, deadtime estimation, and read/write speed tests, are presented in Sections~\ref{sec:data_max}, \ref{sec:data_deadtime} and~\ref{sec:speed_test}.

\subsection{Design and Structure of CdTe-DE Software} \label{sec:struct_de}
The CdTe-DE communicates with each CdTe-DSD and the Formatter via SpaceWire/RMAP, as shown in Fig\,\ref{data_network}.
To ensure reliable communication, we utilized the open-source ``SpaceWire RMAP Library version 2"\cite{SpaceWireRMAPLibrary}, a C++ class library that offers a simple and robust interface for SpaceWire and RMAP transactions\cite{yuasa2008portable}.
RMAP operations are implemented as standard class methods, providing an abstraction that reduces software development cost and complexity while ensuring high reliability.

The CdTe-DE software is composed of multi-threaded C++~programs. It utilizes the StoppableThread class (based on POSIX threads) provided by CxxUtilities, an open-source header-only C++~utility library\cite{CxxUtility}.
One thread is dedicated to generating SpaceWire Timecodes cyclically at 64 Hz, with values ranging from 0 to 63. These Timecodes serve as a master timing reference for synchronizing operations across the CdTe-DE system. The second thread serves as the main control and data acquisition program, and schedules its operations based on the current Timecode value, including control commands and detector readouts (see Appendix\,\ref{sec:design_data} for details).

\begin{figure}[h]
    \begin{center}
    \includegraphics[width=1.0\hsize]{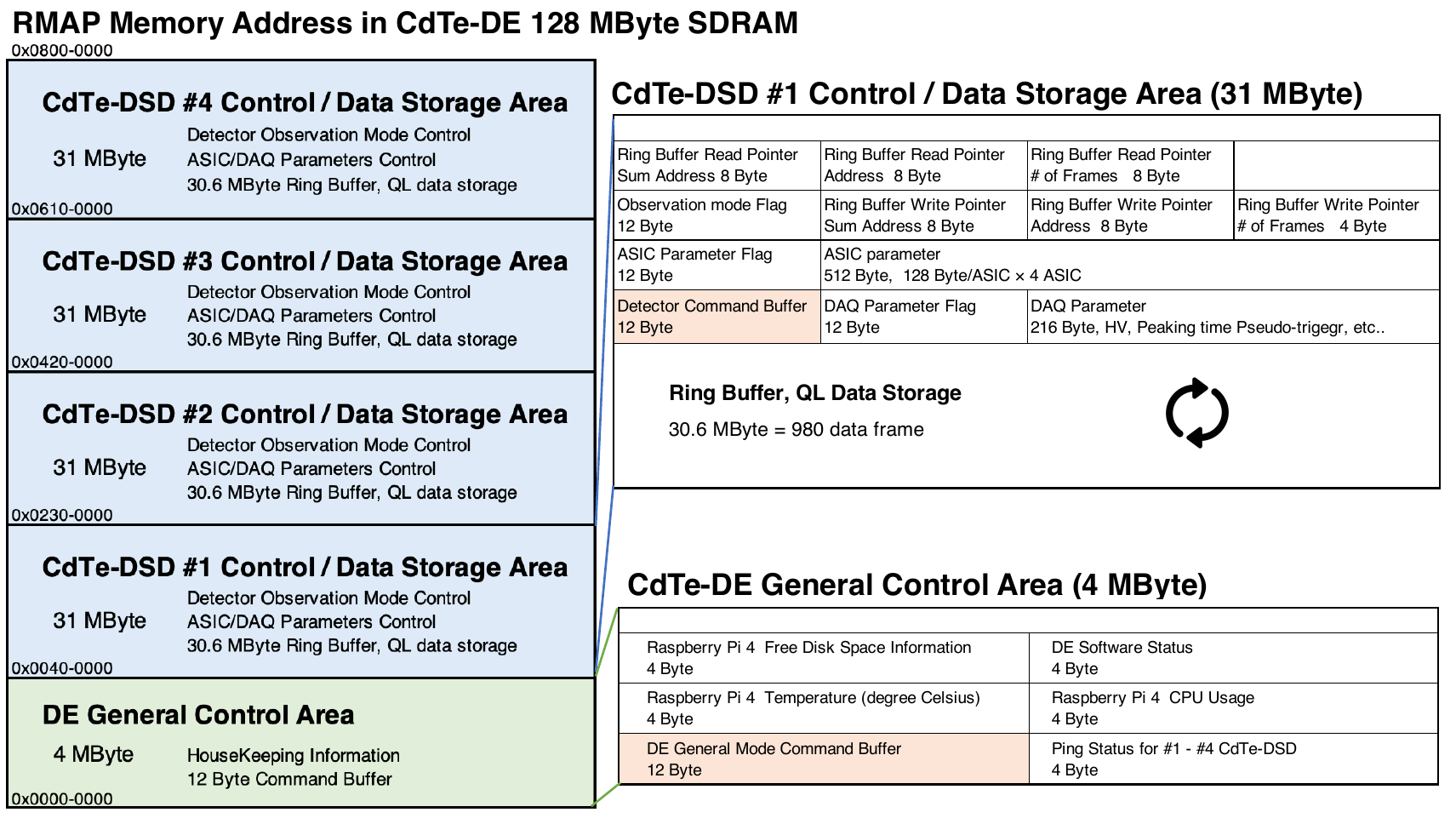}
    \end{center}
    \caption{Memory map of 128 MB SDRAM in the CdTe-DE SPMU-001}
    \label{data_addressmap}
\end{figure}

The CdTe-DE communicates with each CdTe-DSD and the Formatter via SpaceWire/RMAP, as shown in Fig\,\ref{data_network}. Commands and data are transmitted by reading and writing data on the 128 MB SDRAM in the SPMU-001. The SDRAM is mapped as shown in Fig\,\ref{data_addressmap}. The first 4 MB is used as the DE general control area, which stores housekeeping data (e.g., detector status, Raspberry Pi temperature, CPU usage, and disk space) and a 12-byte command buffer from the Formatter to control the CdTe-DE operation mode, which is periodically polled by the CdTe-DE once per second. The remaining 124 MB is divided into four 31 MB regions, one per detector, each designated as a CdTe-DSD control/data storage area. Each region contains a 30.6 MB ring buffer used to store QL data, which the Formatter retrieves and transmits to the ground system for generating spectra and images. The remaining 0.4 MB in each region holds DAQ/ASIC parameters and ring buffer information such as read/write pointers.

\subsection{Event Data Size Optimization} \label{sec:data_max}
For FOXSI-4 and FOXSI-5, the CdTe-DSD readout employs the VATA451.2 ASIC\cite{watanabe2014si}, a member of the VATA series optimized for FOXSI. To enhance data throughput, four ASICs are connected in parallel instead of being daisy-chained, allowing for simultaneous readout. The system supports a sparse readout mode, in which only channels exceeding a user-defined ADC digital threshold ($D_{th}$) are read out, effectively reducing data volume compared to the full-channel readout mode. Temporarily high event rates, such as those during solar flare peaks, are handled using a 64 MB ring-buffer in the SDRAM implemented on the SPMU-001 for each electronics canister.

\begin{figure}[h]
  \begin{center}
  \includegraphics[width=1\hsize]{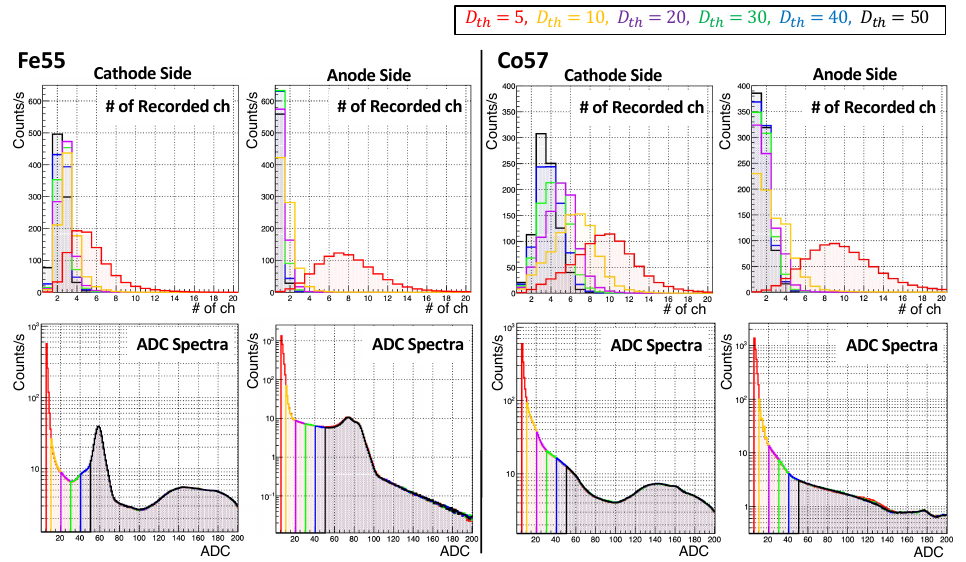}
  \caption{(Upper): Distribution of the number of recorded channels for each $D_{th}$ parameter. (Lower): ADC spectra for all channels for ${}^{55}$Fe (5.9 keV peak) and ${}^{57}$Co (14 keV peak).}
  \label{data_dist_record_ch}
  \end{center}
\end{figure}

The parameter $D_{th}$ plays a critical role in balancing readout efficiency and data quality.
Fig\,\ref{data_dist_record_ch} shows how the distribution of recorded channels varies with the $D_{th}$ setting, based on the measurements under uniform irradiation with X-rays from ${}^{55}$Fe (5.9 keV) and ${}^{57}$Co (14 keV) at -20\,${}^\circ$C (operation temperature).
Increasing $D_{th}$ reduces data volume and increases the number of events that can be packed into each data frame, thereby improving readout speed (see Appendix\,\ref{sec:design_data} for the details of data structure). However, excessively high thresholds may result in the loss of low-energy or peripheral signals. Based on the study of spectral and imaging performance evaluations\cite{nagasawa2025imaging}, a threshold of 1.5 keV is sufficient to preserve key information for energy and position reconstruction. Therefore, the flight configuration adopts $D_{th} = 10$, corresponding to approximately 1.0 keV, which allows around 310 events to be recorded per frame.
\subsection{Deadtime Analysis for Each Electronics Canister} \label{sec:data_deadtime}
To evaluate the system's capability to meet the required acquisition rate of 5000 events/s per detector, we used a high-flux synchrotron X-ray beam at the SPring-8 Synchrotron Radiation Facility in Japan\cite{goto2001construction}. We used the synchrotron X-ray beamline BL20B2, and the CdTe-DSD was irradiated with a 16 keV monochromatic beam (see Section 4 in Ref.~\citenum{nagasawa2025imaging} for the detailed experimental setup).
The deadtime was estimated using the pseudo-trigger method~\cite{kokubun2007orbit}. In this method, pseudo-random triggers are inserted into the data stream at a known rate (10 Hz), and the deadtime is calculated as the ratio between the number of inserted and successfully recorded pseudo-trigger events.
In a dataset with a total exposure of 3960 seconds, 78958 pseudo-trigger events were injected, of which 22060 events were successfully recorded. This yields a livetime fraction of 0.28 and a corresponding deadtime fraction of 0.72.
Excluding pseudo-trigger events, a total of 16470093 events were recorded, corresponding to an average rate of 4159 events/s. From these statistics, the average deadtime per event is estimated as 173~$\mathrm{\mu s}$/event.
Assuming a maximum tolerable deadtime of 90\%, the corresponding maximum event rate is $0.9/173~\mathrm{\mu s} = 5194$ events/s.
This confirms that the required 5000 events/s acquisition rate is achievable with an acceptable level of deadtime.
We note that the absolute photon flux at the detector surface was not independently monitored in this test, because the primary purpose was to characterize the deadtime behavior at a representative high event rate; the detection efficiency of the CdTe-DSD has been characterized separately in Ref.~\citenum{nagasawa2025imaging}.

\subsection{Evaluation of Data throughput from Electronics Canister to CdTe-DE}\label{sec:speed_test}
In nominal flight observation operation, the CdTe-DE reads data from each detector for a duration of 0.468 seconds once every 2 second period, corresponding to 30 out of 128 steps in the 64 Hz Timecode cycle (see Appendix\,\ref{sec:design_data} and Fig\,\ref{data_timecode} for details). 
During this readout window, observation data are read from each electronics canister and saved as binary files on the CdTe-DE’s SD card.
In addition, the most recent frame is written to the CdTe-DE SDRAM for QL data, which is then extracted by the Formatter.

To evaluate the data throughput from each canister to the CdTe-DE, we measured the required processing time. The dominant bottleneck in this process is the SpaceWire link speed between each canister and the CdTe-DE.
In FOXSI-4, the link speed was configured at 10 Mbps, under which the CdTe-DE could process approximately 4 frames/s/canister, corresponding to $\sim$1240 events/s/canister. For FOXSI-5, the link speed is increased to 50 Mbps, allowing a throughput of approximately 12.5 frames/s/canister, or $\sim$3875 events/s/canister.
Although this rate does not reach the maximum data acquisition capability of 5000 events/s for each canister, each electronics canister contains a 64 MB ring buffer. This buffer enables temporary storage of high-rate data, which can later be retrieved by the CdTe-DE during periods of lower count rates or after the end of an observation.

\section{CMOS Data Electronics} \label{sec:cmos_daq}
The CMOS Data Electronics system for FOXSI-4 and FOXSI-5 was mainly developed and provided by NAOJ.
While CMOS sensors were first introduced in FOXSI-3 \cite{NARUKAGE2020162974}, the readout system lacked the capability to take  all the data output from the CMOS sensor. This was due to the performance of the CPU part of the system-on-chip. Additionally, the system did not include the telemetry support for QL data and command-based control of the CMOS sensors.
This was because FOXSI-3 reused the existing FPGA-based readout architecture, which had been designed only for the hard X-ray detectors (Si-DSD and CdTe-DSD) used in FOXSI-1 and FOXSI-2. In FOXSI-3, after the soft X-ray CMOS camera was powered on, it was operated by the camera system timer as a standalone system.

For the solar flare observation of FOXSI-4 and FOXSI-5, real-time control and monitoring of the CMOS sensors became essential, especially to support flexible observation modes, especially exposure time adjustment, depending on flare intensity. 
To meet these requirements, the CMOS data electronics system was designed with the following key capabilities:
\begin{itemize}
    \item Onboard storage of all acquired photon-counting data to an SSD;
    \item Downlink of QL images and a portion of photon-counting data for real-time monitoring and solar pointing assessment;
    \item Command-based control of the CMOS sensors from the Formatter, including sensor exposure start/stop and dynamic exposure time adjustment.
\end{itemize}
To implement this functionality, a new data acquisition architecture was developed based on the Zynq FPGA readout system used in FOXSI-3 \cite{ISHIKAWA2018191}.
This upgraded system consists of a newly developed FPGA board, SPMU-002, which incorporates a Zynq UltraScale+ MPSoC (Section \ref{sec:spmu002}).
It also includes updated FPGA logic that enables optimized sensor exposure modes (Section \ref{sec:cmos_exp}), along with Linux-based software that performs sensor control, manages onboard data storage, and generates QL data for downlink via the telemetry system (Section \ref{sec:cmos_soft}).

\subsection{SPMU-002 FPGA Board}\label{sec:spmu002}
\begin{figure}[h]
    \begin{center}
    \includegraphics[width=1.0\hsize]{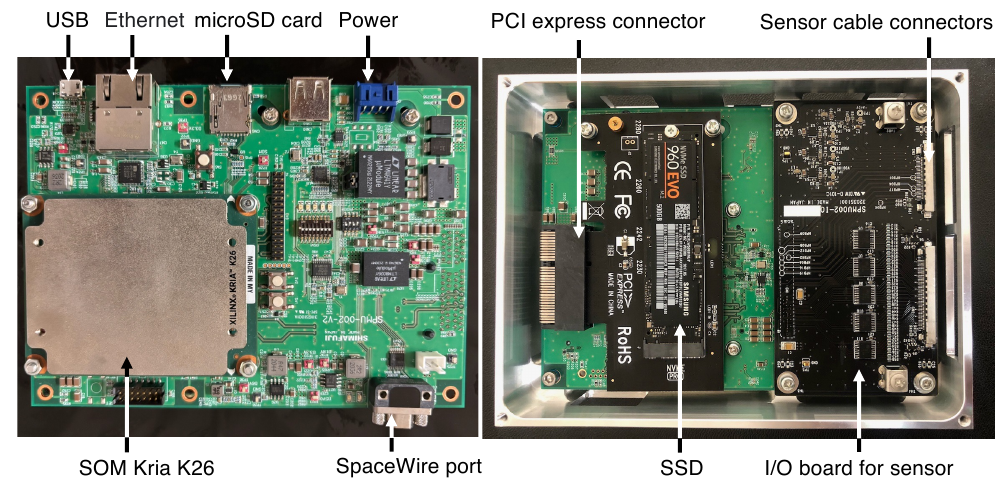}
    \end{center}
    \caption{Pictures of the SPMU-002 FPGA board. (Left) Front side of the board. SpaceWire port is not mounted in this picture. (Right) Back side of the board, enclosed in the electronix box with the top lid open.}
    \label{fig:SPMU002_pic}
\end{figure}

The SPMU-002 is an FPGA-based data acquisition board jointly developed by Shimafuji Electric Incorporated, Kavli IPMU, and NAOJ to handle high-speed data readout from CMOS sensors and enable SpaceWire communication (see Fig~\ref{fig:SPMU002_pic}). At the core of the board is an AMD Kria K26 System-on Module (SOM), which integrates an AMD Zynq UltraScale+ MPSoC. This architecture combines a quad-core Arm Cortex-A53 application processor in the Processing System (PS) with high-performance Programmable Logic (PL), allowing the board to coordinate Linux-based control and FPGA-based real-time processing within a single module.
Leveraging this architecture, SPMU-002 can process up to approximately 375 MB/s of incoming data from the CMOS sensors in real time, while also providing robust SpaceWire communication with external subsystems.

\begin{figure}[h]
    \begin{center}
    \includegraphics[width=1\hsize]{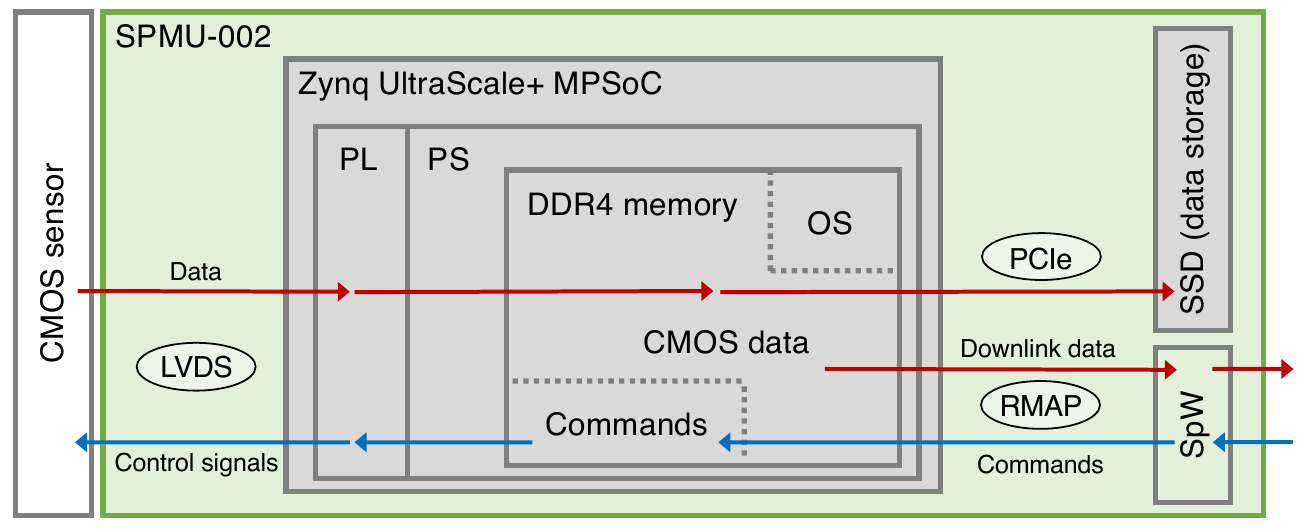}
    \end{center}
    \caption{Schematic overview of the command and data readout architecture implemented for FOXSI-4 and FOXSI-5.}
    \label{fig:SPMU002_diagram}
\end{figure}

Fig~\ref{fig:SPMU002_diagram} shows a schematic overview of the command and data readout architecture implemented for FOXSI-4 and FOXSI-5 using the SPMU-002 board.
Sensor data are received via an LVDS bus into the PL side of the Zynq device. These data are then passed to the Linux system operating on the PS side and written as files to an onboard SSD via a PCI Express interface. The Linux system, along with application software and the FPGA boot image, is preloaded onto a microSD card and automatically loaded at system startup.

During ground-based testing, the system can be accessed from an external PC via serial communication (USB connector) or over a network using SSH (Ethernet connector).
Software can be updated on an external PC and transferred to the microSD card via SSH, and compiled directly on the Linux system.
This architecture significantly improves development efficiency.
In addition, software debugging and functional testing can be performed interactively through serial or SSH communication, allowing for efficient testing and verification from an external PC.

During flight, telemetry is sent via the SpaceWire interface. Using the SpaceWire/RMAP protocol, data stored in the PS-side DDR4-SDRAM can be retrieved and transmitted to the Formatter.
Commands from the Formatter are also received via SpaceWire by writing to designated memory regions in the DDR4-SDRAM, which are continuously monitored by the Linux software running on the PS side.

\subsection{Sensor Exposure and Readout Optimization}\label{sec:cmos_exp}
The CMOS system supports two exposure modes: Test mode and Flight mode. Both modes are implemented at the FPGA level.
Test mode follows a conventional full-frame exposure method, in which all 2048$\times$2048 pixels of the sensor are exposed simultaneously for a fixed integration time to generate a single image. This standard approach is typically used for calibration and functionality checks on the ground.

\begin{figure}[h]
    \begin{center}
    \includegraphics[width=1.0\hsize]{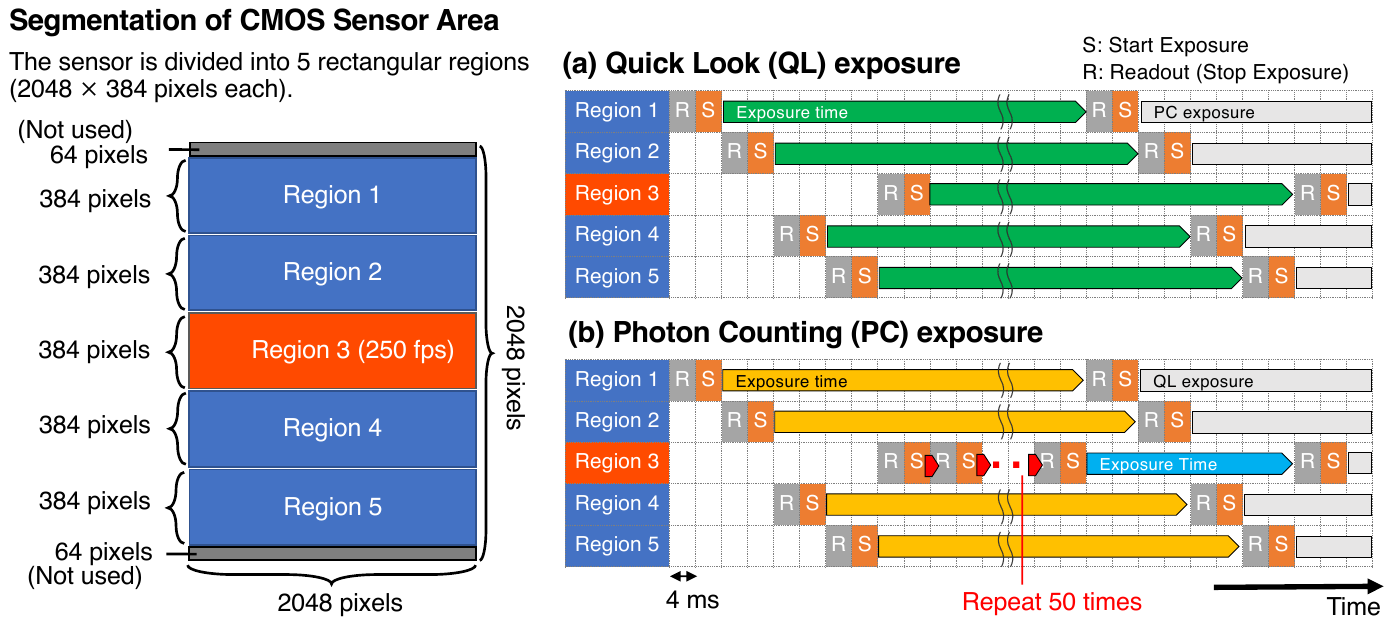}
    \end{center}
    \caption{(Left) Conceptual diagram of CMOS sensor region segmentation. The top and bottom 64 lines are unused, and the remaining 1920 lines are divided into five regions of 384 lines each. Continuous high-speed exposure at 250 fps is performed in the central Region 3. (Right) Conceptual timing diagram of the exposure sequence in Flight mode. The horizontal axis represents time, where ``R" denotes readout (end of exposure) and ``S" denotes the start of exposure. Each R or S operation takes 4 ms per region. In flight mode, (a) quick-look exposures and (b) photon-counting exposures are  executed in an alternating sequence.}
    \label{fig:SPMU002_readout}
\end{figure}

In contrast, Flight mode is optimized specifically for solar flare observations during flights.
Fig~\ref{fig:SPMU002_readout} shows the concept of the exposure sequence in Flight mode.
In this mode, the sensor area is divided into five rectangular regions (Region 1 through Region 5), each consisting of 2048$\times$384 pixels.
The system alternates between two exposure patterns: Quick-Look (QL) exposure and Photon-Counting (PC) exposure.
During a QL exposure, the system initiates exposure sequentially in Regions 1, 2, 4, 5, and 3.
After a predefined integration time, exposure ends in the same order. Thus, all regions share the same exposure time.
QL data are used to verify the field of view and support pointing operations in flight. Each QL frame includes a header that stores metadata such as frame ID and exposure parameters.
In contrast, during a PC exposure, Regions 1, 2, 4, and 5 begin exposure first. Region 3 then undergoes 50 cycles of short exposures, each lasting 4 ms.
After this burst sequence, Region 3 starts to be exposed again, and exposure ends across all regions in the order of 1, 2, 4, 5, and 3.
This scheme can perform photon-counting observations for a wide solar area of both flaring and its surrounding regions.
The short (4 ms) exposures in the central Region 3, where the intense flare emission is expected, enable photon-counting with high temporal resolution, while effectively longer (200 ms) exposures are used to perform the photon-counting for the surrounding regions where emission is expected to be less intense than the flaring region.

\subsection{Software Design}\label{sec:cmos_soft}
The onboard software running on the Linux system of the SPMU-002 board is responsible for executing sensor control commands, managing sensor data storage, and preparing downlink telemetry.
It provides the following core functionalities:
\begin{itemize}
    \item Command handling for sensor control: All control commands, including exposure start and stop, as well as parameter adjustments, are issued from the Formatter. The software interprets incoming commands and executes the corresponding actions;
    \item Sensor data storage: data from the CMOS sensor are saved to the onboard SSD;
    \item HK data management:
    Key operational parameters such as camera operation status, data storage space, CPU load average, FPGA temperature and exposure settings are monitored, recorded, and stored both in memory for downlink and on the SSD;
    \item Preparation of telemetry science data:
    A portion of the sensor data is extracted, formatted, and stored in designated memory regions for downlink during flight.
\end{itemize}

Commands from the Formatter and buffering of QL data are handled via a dedicated 128 MB memory region allocated within the CMOS system. This memory is accessible from the Formatter using the SpaceWire/RMAP protocol.
Within this memory, 
416 bytes are allocated for receiving commands and reporting their execution status. An additional 352 bytes are reserved for HK data storage, while the remaining area is used to store QL science data for downlink.

All commands necessary for controlling the CMOS system are defined in a command table. Each command is defined as a fixed 4-byte word and executed by writing it to its dedicated memory address.
For example, writing a command to address 0x20 stops the ongoing exposure, resets all sensor settings to their default values, and then restarts exposure.
This operation is used when abnormal outputs are detected from the sensor due to desynchronized LVDS timing.
It also serves as a full reset of the exposure parameters.
To prevent unintended execution, emergency commands such as system reboot are implemented with a double-command mechanism. These commands are only carried out if a dedicated flag in a separate memory region is first set, followed by the reception of the corresponding execution command.

Each CMOS sensor generates data at approximately 375 MB/s during continuous exposure.
However, the available downlink bandwidth is limited to 640 kB/s per sensor, and the Formatter reads data once per second, making it infeasible to transmit all acquired data during flight.
Therefore, all data are stored on an onboard SSD for post-flight recovery, and only selected data necessary for real-time monitoring during flight are downlinked, including system status and command execution, QL images for telescope pointing, and PC data for exposure time adjustment if needed.
In order to fit these selected data within the available bandwidth, low-volume status and HK data (512 bytes in total) are transmitted every second.
In contrast, QL and PC sensor data are alternately downlinked every second, resulting in each type being transmitted every two seconds.
The QL data consist of binned images of the full sensor area, reduced to 481 kB per frame, while the PC data retain raw pixel values but are spatially limited to the central 768$\times$384 pixel region, yielding 577 kB per frame.
All downlink datasets are stored in predefined memory regions, from which the Formatter retrieves them for transmission to the ground via the telemetry.

\subsection{Ground-based Performance Evaluation Test}\label{sec:cmos_test}

Photon counting with the CMOS system is achieved by acquiring image data under conditions where photons enter the detector sparsely during each exposure, and then extracting individual photon events through post-processing.
The photon incidence rate per exposure is adjusted using an attenuation filter to maintain an appropriate level for photon counting.
If the actual rate exceeds the designed value, multiple photons may be detected within the same pixel during a single exposure, which prevents photon counting.
In cases of exceptionally high incidence rates, for example, images of the corona or flares will be acquired, similar to a typical imaging observation.
Importantly, regardless of the incidence rate, the data acquisition rate of the CMOS system remains constant.
FOXSI-4 achieves a high-speed exposure rate of 250 frames per second for a 384$\times$2048 pixel region, which corresponds to a data rate of approximately 375 MB/s.
The feasibility of this rate was confirmed in ground-based camera performance evaluation tests by checking the exposure information embedded in each file and verifying that all data were written to the SSD without loss.
These tests were performed under various conditions, including operation in air, in vacuum, during cooling, and under X-ray irradiation.

\section{Quad-Timepix3 Data Electronics} \label{sec:timepix_daq}
The Quad-Timepix3 system provided by SSL/UC Berkeley for the FOXSI-4 payload marked the first use of a Timepix3 ASIC for solar flare observation. While the Timepix ASIC family has a long heritage at SSL/UC Berkeley, including prior use with a microchannel plate detector operating at kilohertz frame rates\cite{Vallerga2011}, and has seen extensive deployment in synchrotron radiation experiments\cite{McCarter23, Tremsin21}, adapting this system for a sounding rocket flight presented unique engineering challenges.
These included minimizing size and power consumption, improving environmental robustness, and integrating with the FOXSI-4 flight and telemetry systems.

The data electronics suite comprises an AMD Kintex-7 FPGA on a Genesys 2 development board and a Raspberry Pi 3B+ microcomputer. These were connected via Ethernet, augmented with a USB 3.2 Gen1 to RJ45 Gigabit Ethernet dongle. This setup enabled support for 9000-byte jumbo frames, a legacy requirement inherited from prior synchrotron-based deployments of the same FPGA architecture.
The FPGA board handled onboard science data packetization, while the Raspberry Pi supported command interfacing from the FOXSI-4 Formatter and telemetry generation for real-time downlink and health monitoring (QL) during integration and flight.
The addition of the Gigabit Ethernet dongle ensured compatibility with the FOXSI-4 system, which did not support the native 10-Gigabit Ethernet (10 GbE) interface used in the original synchrotron deployment. The 10 GbE interface was initially selected to meet strict timing requirements and minimize development time by leveraging the same FPGA architecture used in a 100 GbE board previously developed for high-throughput applications.

% In total, the Timepix3 Data Electronics system was responsible for:
Having integrated these hardware and interface components, the Timepix3 Data Electronics system was designed as a compact and autonomous readout unit, combining hardware reuse, streamlined software design, and tradeoff studies addressing thermal management, telemetry constraints, and data handling.
These design optimizations aimed to balance performance and reliability within the thermal and operational limitations of a sounding rocket flight, while also supporting the following key operational functionality required during flight:
\begin{itemize}
    \item Interfacing with and processing event data from four Timepix3 ASICs (constituting one quad detector module);
    \item Writing raw measurement packets in package capture (PCAP) format to local Raspberry Pi storage;
    \item Generating and formatting telemetry packets for real-time monitoring and flight downlink.
\end{itemize}

\subsection{Operational Modes, Commanding, and Quick-Look System}

The Timepix3 system operated in a single autonomous mode, optimized for continuous solar observation. 
After the ASICs and FPGA were powered on, the Raspberry Pi booted and applied predefined configuration settings to the ASICs.
% Upon power-up, the Raspberry Pi initiated a boot sequence that loaded predefined configurations onto the ASIC. Upon FPGA power-up, the ASICs were also powered. 
Once configured, the system entered its nominal operating state and initiated the fixed measurement loop, during which the shutter door was opened.

During science operations, data flowed from the Timepix3 detector to the FPGA, where event packets were buffered and formatted. The FPGA then transmitted the data via Ethernet to the Raspberry Pi, which simultaneously forwarded it to the Formatter for downlink and saved it to local onboard storage for post-flight retrieval.
Each photon-counting measurement consisted of a fixed-duration exposure, acquired at a one-second cadence. The system (ASICs and FPGA) accumulated photon events into Ethernet jumbo frames (up to 9000 bytes each), which included pixel event data such as spatial coordinates, Time over Threshold (ToT, corresponding to energy), and timing information, as well as auxiliary metadata required to reconstruct UTC timestamps and perform full event calibration.
% Data flowed from the Timepix3 detector to the FPGA during science operations, where event packets were buffered and formatted. The FPGA then passed this data via Ethernet to the Raspberry Pi, which concurrently:
% \begin{itemize}
%     \item Forwarded it directly to the Formatter for downlink, and
%     \item Saved it locally to onboard storage for post-flight retrieval.
% \end{itemize}
% Each measurement consisted of a fixed-duration exposure, acquired on a one-second cadence. The system accumulated science data into Ethernet jumbo frames—each with a maximum size of 9000 bytes—before transmission. These frames included pixel event data (e.g., spatial coordinates, Time over Threshold (ToT), and timing information) and auxiliary metadata required to reconstruct UTC timestamps and perform full event calibration.
% Each photon-counting measurement was performed with a fixed-duration integration exposure, acquired at a one-second cadence. The system (ASICs and FPGA) accumulated photon data into 9000-byte packets, which were subsequently transmitted to the microcomputer for storage and dissemination. Each photon data packet included pixel event data (e.g., spatial coordinates, Time over Threshold (ToT), and timing information) and auxiliary metadata required to reconstruct UTC timestamps and perform full event calibration.

Since the Timepix3 system operated in a fixed autonomous mode and did not require dynamic retasking during flight, it was designed without the need for in-flight command capability. However, a limited command and telemetry interface was implemented, allowing the Formatter to:
% The Timepix3 system did not require dynamic retasking during flight and therefore operated without mission-critical command capability. However, a limited command interface was implemented, allowing the Formatter to:
\begin{itemize}
    \item Request internal flags and housekeeping information for system status monitoring;
    \item Issue system resets and delete onboard data storage if memory capacity is reached during preflight testing or while on the launch rails;
    \item Trigger basic flux and energy measurements (in ASIC-native units) as a coarse confirmation of solar pointing, although the Timepix3 was not involved in real-time pointing decisions.
\end{itemize}
To support real-time system monitoring, two types of telemetry packets were transmitted. The first was housekeeping telemetry, which provided basic boot status and software health monitoring.
Using these status flags and a GSE display, this telemetry enabled the ground team to identify anomalous values for immediate review and confirm that the system remained in science mode following boot.
In the case of a critical fault, the Formatter could remotely power-cycle the Timepix3 subsystem to restore nominal operation without disrupting other onboard instruments.
The second was rate-based telemetry, which provided live insight into detector performance during flight. This system helped confirm proper operation and enabled ground teams to interpret detector activity in context by reporting high-level science metrics, including the mean ToT per second and overall event rates.
These metrics were chosen because they provide immediate feedback on the observation condition. 
For example, an increase in ToT or event rate could be used to confirm that the telescope was pointed at the Sun without requiring full image reconstruction.
This approach reduced telemetry load while allowing science teams to verify system performance in real time and reserved bandwidth for the downlink of critical image and spectroscopic data from other instruments.

% Two types of telemetry packets were sent and displayed live:
% \begin{itemize}
%     \item \textbf{Housekeeping (HK)} telemetry included software flags, boot status, and system health indicators.
%     \item \textbf{Rate-based telemetry} reported aggregate science metrics, including mean ToT per second and event rates (flux).
% \end{itemize}

\subsection{Optimization for Solar Flare Observation}
To optimize the Timepix3 system for high-rate solar flare observations, a trade-off study was conducted to balance thermal constraints, data throughput, and telemetry efficiency. 
% Below, we summarize key considerations that shaped the design.
The following subsections summarize key considerations that informed the system design.

\subsubsection{Deadtime and Thermal Constraints}
Although the Timepix3 ASIC is capable of timing resolutions down to 1.56 ns with near-zero deadtime in data-driven mode\cite{Pitters19}, these capabilities could not be fully exploited during FOXSI-4 due to thermal limitations. 
Operating the ASICs at such high throughput would have significantly increased power consumption\cite{Burian19}, generating thermal loads beyond the capacity of the payload’s cooling system. 
This system used liquid nitrogen delivered in gaseous form, which was directed onto the detector plate and provided only limited cooling efficiency.
Furthermore, additional heat from the Timepix system would have interfered with the cryogenic environment of the focal plane, which was measured to be as low as –14\textdegree C on the launch rails by Timepix and was critical to maintaining detector performance during flight.
% The liquid nitrogen used for cooling was delivered in gaseous form and directed onto the detector plate.
% In particular, additional heat from the Timepix system would have interfered with the focal plane’s cryogenic environment, which was measured to be as low as –14\textdegree C on the rails by Timepix and was critical for maintaining detector performance during flight.

To accommodate these thermal constraints, system parameters were tuned to reduce power draw at the expense of some deadtime.
Even with reduced power settings, ground testing with sealed sources demonstrated effective deadtime-free operation at flux levels consistent with in-flight conditions.
This is consistent with Timepix3 performance reports from CERN, which estimate minimum deadtime contributions on the order of 375 ns per packet transfer, and support hit rates up to 40 Mhits/s/cm${}^2$ with full clock speeds\cite{Gaspari14}.

\subsubsection{Data Structure and Throughput}
Science data were transmitted from the FPGA to the Raspberry Pi in UDP packets configured with the jumbo frame size of 9000 bytes.
Each packet began with a 42-byte header composed of standard Ethernet, IP, and UDP fields, followed by 6 bytes of internal metadata including the ASIC chip ID, the elapsed measurement time, and a packet index used for ordering during reconstruction.
The remainder of the packet contained raw ASIC data, comprising pixel coordinates, ToT, and timing information necessary for converting event times into UTC.
In parallel, a separate service packet of 168 bytes was also periodically generated.
This packet contained housekeeping information such as voltages and currents from the FPGA and ASICs, as well as temperature readings from both the FPGA and the main system board.
Together, the science and service packets provided the necessary context for accurate post-flight data reconstruction and calibration.
% Science data were transmitted from the FPGA to the Raspberry Pi in jumbo UDP packets, each 9000 bytes in size. These packets followed the structure:
% \begin{itemize}
%     \item First 42 bytes: Standard Ethernet, IP, and UDP headers;
%     \item Next 6 bytes: Packet metadata, including ASIC chip ID, elapsed measurement time, and a packet index for reconstruction ordering;
%     \item Remaining payload: Raw ASIC data, including pixel coordinates, time over threshold (ToT), and timing information necessary for later reconstruction in UTC.
% \end{itemize}
% In addition, a separate 168-byte ``service packet" was periodically generated, containing housekeeping (HK) information such as:
% \begin{itemize}
%     \item FPGA and ASIC voltages and currents,
%     \item Onboard temperatures from the FPGA and system board.
% \end{itemize}
% The science and service packets provided the context required for post-flight data reconstruction and calibration.

\subsubsection{Telemetry Interface and Minimal Processing}

Unlike other subsystems in the FOXSI-4 payload, the Timepix3 electronics did not use the SpaceWire/RMAP protocol. Instead, a UART interface was implemented between the Raspberry Pi and the Formatter. Operating at 9600 baud, this serial link was sufficient for low-rate telemetry exchange, as the system was not used for active solar pointing decisions.

All telemetry was produced by Python scripts running on the Raspberry Pi. The \texttt{tshark}\cite{tshark} module was used to capture Ethernet traffic directly from the FPGA and write it to local storage of a 1 TB SanDisk SD card. To minimize onboard processing demands, telemetry was intentionally kept simple: all packets were 27 bytes or smaller and transmitted only when polled by the Formatter. 
This design prioritized reliable data capture and thermal stability over complex onboard processing or real-time analysis.

\subsubsection{FOXSI-5 Quad-Timepix3 Telemetry Updates}
All telemetry functionality from the FOXSI-4 flight is retained for FOXSI-5, with data requested and downlinked at the same expected cadence. In addition to this baseline telemetry, two new packet types have been introduced to enhance system monitoring and streamline pre-flight alignment procedures.
The FOXSI-4 Timepix3 failure was ultimately traced to an initialization issue where powering the FPGA back on too soon left residual charge, disrupting communication with the ASICs and leaving them powered but unconfigured. Consequently, flight data from the ASICs was empty, as configuration had failed before the shutter door opened. To prevent recurrence, FOXSI-5 introduces two telemetry additions: a ``Timepix3 image packet", which confirms that the ASICs are producing photon data, and a ``PCAP-status packet", which verifies that the FPGA is generating appropriately sized data files rather than the empty or anomalous files observed during FOXSI-4.

The Timepix3 image packet contains the most recent 360 photons captured by the detector. Through the GSE, these photon lists are formatted into a two-dimensional image and displayed in real-time.
% The display can be manually cleared and re-aggregated by the user.
This feature was developed to simplify the laser alignment process: during FOXSI-4, each adjustment required offloading data from the Timepix system, processing it on a separate computer, and then evaluating the resulting image. By enabling direct transmission of image data to the GSE, alignment procedures involving Timepix-centered optics can be performed more efficiently. Although the infrastructure supports in-flight downlinking of image packets, there are currently no plans to implement this, as the data is not used for pointing decisions, consistent with the approach taken in FOXSI-4.

The PCAP status packet reports the sizes of the four most recently saved PCAP files on the microcontroller. During previous testing, detector startup issues were often correlated with anomalously large PCAP files, which increased in size from the typical few KB to several hundred MB and contained primarily electronic noise. The inclusion of this packet provides a real-time diagnostic metric: large or erratic PCAP sizes can indicate improper system boot and may be used to inform a decision to initiate a power cycle.

These diagnostics work in tandem with the FOXSI-5 hardware mitigation, in which a bleed resistor that ensures the FPGA fully discharges during power-off, to provide both preventive and real-time confirmation of proper startup. Bench testing, including cold-temperature operation and rapid power-cycle scenarios, demonstrates that the new packets reliably detect improper initialization and validate when the detector has reached a proper data-taking state.

In addition to the new packets, the flag system used to monitor the health of the Timepix system has been updated to include a ``process check" flag. This flag is triggered if any critical parallel process unexpectedly terminates, and the event is logged with a timestamp to aid in post-flight diagnostics.

% The Timepix3 data electronics system developed for FOXSI-4 marked a novel effort to adapt high-resolution ASIC-based detectors for solar flare observation on a suborbital platform. 
% The system was designed as a compact and autonomous readout unit through a combination of hardware reuse, streamlined software design, and tradeoff studies addressing thermal management, telemetry constraints, and data handling.
% These design optimizations aimed to balance performance and reliability within the thermal and operational limitations of a sounding rocket flight. 

% Through hardware reuse, streamlined software design, and practical tradeoffs—such as thermal management constraints, simplified telemetry, and efficient data handling—the system was designed to operate as a compact, autonomous readout unit under the environmental and operational conditions of flight—these design choices aimed to balance performance with reliability in a constrained flight environment.

\section{Ground Support Equipment (GSE)} \label{sec:gse}
The ground support equipment (GSE) used for the FOXSI-4 flight\cite{GSE_soft} is separated into three parts: commanding (Section~\ref{sec:gse-commanding}), data logging (Section~\ref{sec:gse-data-logging}), and data display (Section~\ref{sec:gse-data display}). These three aspects form the pipeline for the display of FOXSI-4 observations taken in both real time and afterwards.

    % \begin{figure}[h!]
    %     \centering
    %     \includegraphics[width=0.8\textwidth]{flight_gse_screenshot.png} 
    %     \caption{The GSE visual display (left panel) and commanding (right panel) windows.}
    %     \label{fig:gse-windows}
    % \end{figure}

%     \begin{figure}
%     \centering
%     \begin{minipage}{0.7\textwidth}
%         \centering
%         \includegraphics[width=1.0\textwidth]{flight_gse_screenshot.png} % first figure itself
%         \caption{GSE display software during FOXSI-4 flight. The top row shows the integrated flight data from each CdTe detector, the middle row shows the display for each CMOS detector, and the bottom left window shows information from the Timepix system. The bottom center panel gives access to controls which rotate and restart the integration of the images. The bottom right window will display any whole-system log messages.}
%         \label{fig:gse-windows-display}
%     \end{minipage}\hfill
%     \begin{minipage}{0.3\textwidth}
%         \centering
%         \includegraphics[width=1.0\textwidth]{gse_uplink} % second figure itself
%         \caption{GSE command uplink window. The system to be commanded is chosen on the left with the system-specific command being selected in the right list. The command window also displays system health information (e.g., communication status and the current).}
%         \label{fig:gse-windows-command}
%     \end{minipage}
% \end{figure}
\begin{figure}[h]
  \begin{center}
  \includegraphics[width=1\hsize]{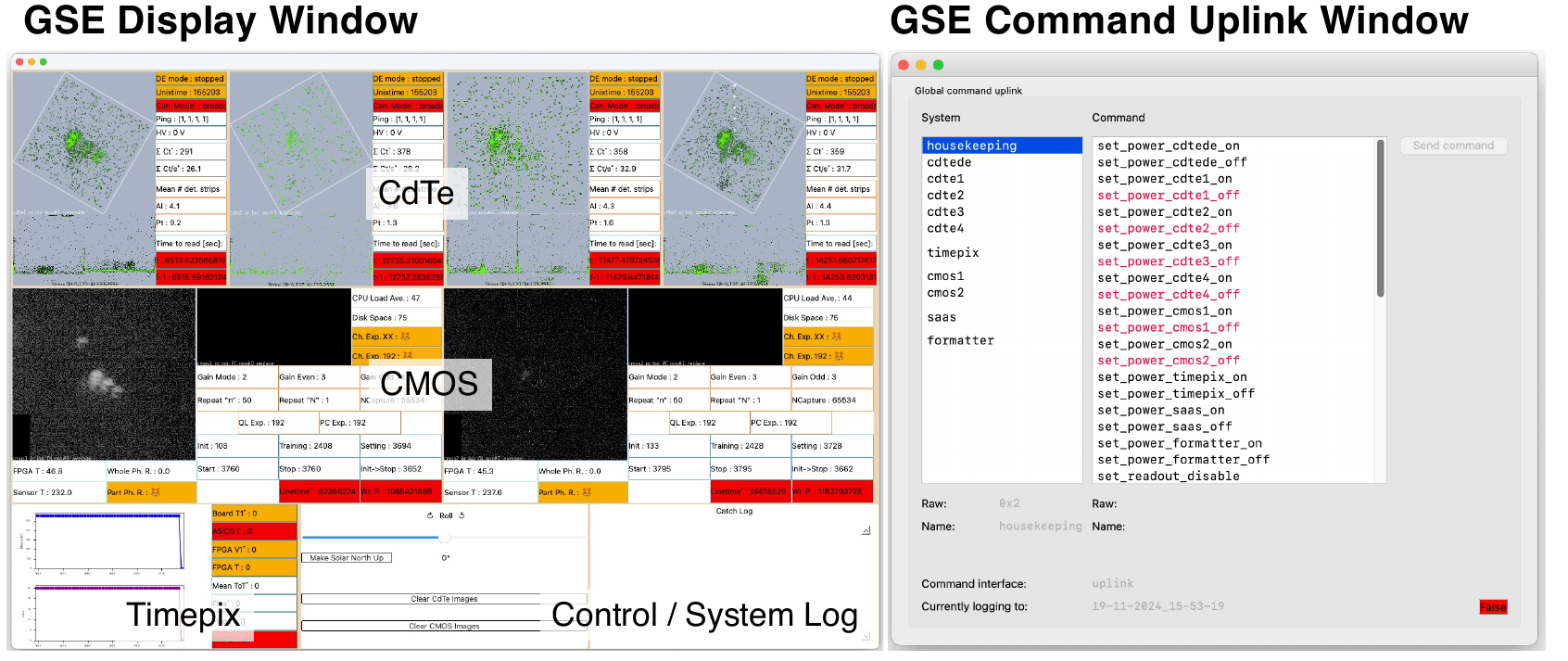}
  \caption{(Left): GSE display software during FOXSI-4 flight. The top row shows the integrated flight data from each CdTe detector, the middle row shows the display for each CMOS sensor, and the bottom left window shows information from the Timepix system. The bottom center panel gives access to controls which rotate and restart the integration of the images. The bottom right window will display any whole-system log messages. (Right): GSE command uplink window. The system to be commanded is chosen on the left with the system-specific command being selected in the right list. The command window also displays system housekeeping information (e.g., communication status and the current).}
  \label{fig:gse_unit}
  \end{center}
\end{figure}

Fig~\ref{fig:gse_unit} left shows the detector data display, visualizing the telemetry data observed during the full FOXSI-4 flight from the seven detectors. Information from the CdTe system is shown at the top of the display, CMOS is shown in the middle, and Timepix is shown in the bottom left. The bottom center displays controls for the QL images, such as rotation angle and restarting image integration. The bottom right panel is the ``Catch Log'' panel, which is reserved for general, whole-system log messages.
Information received from all four CdTe detectors is organized to show QL images (rotated with respect to focal plane position), spectrograms, and status information such as high-voltage bias and count rates. Similarly, telemetry from the CMOS system includes QL images (larger grayscale panels) and status information, such as on-board clock values and observation status. The smaller, rectangular panels show emission recorded by the photon-counting region of the CMOS sensor. The Timepix section shows time profiles of mean ToT and flux on the left and status information on the right.
The command uplink box (Fig~\ref{fig:gse_unit} right) was used by operators during pre-launch activities to power detector systems on and off, configure them, begin data acquisitions, and test required functionality for use in flight. 

\subsection{Commanding} \label{sec:gse-commanding}
Uplink commands from the GSE to the FOXSI payload system were used to control essential functions, including powering detector systems on and off, configuring detector parameters, and starting or stopping data collection. These operations were routinely performed during pre-flight testing on the ground.
In flight, several critical command capabilities were also required. These included raising the detector bias voltage only after reaching a specific altitude to avoid arcing, adjusting detector thresholds in response to high flux levels, and shutting down systems safely at the end of the observation. All of these actions were executed via the GSE command uplink interface, with detector teams monitoring the GSE display for command feedback.

% There are several uses for uplink commands from GSE to FOXSI payload system:
% \begin{itemize}
%     \item power detector systems on and off,
%     \item set configuration of detector systems,
%     \item start and stop data collection.
% \end{itemize}
% These commands were all routinely used on the ground during pre-flight tests. It was also necessary to send a few critical commands in flight---to raise the detector bias voltage only above a particular altitude (to avoid arcing), to modify detector thresholds in case of excessive flux, and to shut down systems appropriately at the end of the observation. The GSE command uplink window allowed an operator to perform all these activities, with feedback from detector experts in the room.

The command uplink window encodes the onboard system and command pair selected by the operator as a two-byte value. That raw command value is uplinked to the FOXSI payload over a radio link, via a UART interface. Onboard, the uplink command is received by the Formatter and sorted into the appropriate command queue for the specified onboard system.

\subsection{Data Logging} \label{sec:gse-data-logging}
The GSE software is responsible for reliable logging of all received telemetry to disk. The telemetry data are logged in a form that is acceptable for live display, but without any additional processing. The recorded data products are raw, binary logs of whatever housekeeping data, event lists, or QL images were handed to the Formatter by onboard detector systems. This design approach was chosen in order to present the Formatter-GSE interface as a transparent pipe for detector data. This approach enables detector data parsing software to ingest telemetry data just as it would ingest raw detector data directly, with no Formatter or GSE inline.

All downlink data products are fixed-size binary data, so a dedicated log file is used to store each telemetry data type for each onboard system (see Section \ref{sec:formatter-sw-downlink}). Data is downlinked in small \textit{packets} which must be reassembled into complete \textit{frames}. Only complete frames should be written to disk, not packets, so that the GSE display functionality is able to parse frame data without error. To that end, packets are inserted into a buffer until all packets for that frame are populated, at which point the frame is written to disk. Algorithm \ref{alg:gse-log} describes, in pseudocode, the procedure used to populate the packet buffer for each system and data type, and to write that buffer to disk. Note that if a packet is dropped, the incomplete frame containing the dropped packet will still be written to disk when the subsequent frame arrives.

    \begin{algorithm}
        \caption{GSE logging logic for a \texttt{packet} with given \texttt{system} and \texttt{datatype}.}
        \label{alg:gse-log}
        \begin{algorithmic}[1]
            \Procedure{log\_packet}{\texttt{packet}, \texttt{buffer\_list}}
                \State \texttt{buffer} $\gets$ \texttt{buffer\_list.lookup(packet.system, packet.datatype)}
                \If{\texttt{buffer.contains(packet.index)}}
                    \texttt{buffer.write()} \Comment{save the buffer to disk}
                \Else{}
                    \texttt{buffer.insert(packet)}  \Comment{insert the packet into the buffer}
                \EndIf
            \EndProcedure
        \end{algorithmic}
    \end{algorithm}

\subsection{Data Display} \label{sec:gse-data display}
The data display (Fig~\ref{fig:gse_unit} left) is a crucial tool during flight; therefore, it must be reliable, fast, and easily interpreted. The display was coded exclusively in Python for readability and maintainability while providing a versatile environment that will work on a variety of systems. The design was carefully created in collaboration with the individual detector teams to support any actions or commanding during flight.

\begin{figure}[h!]
    \centering
    \includegraphics[width=0.8\textwidth]{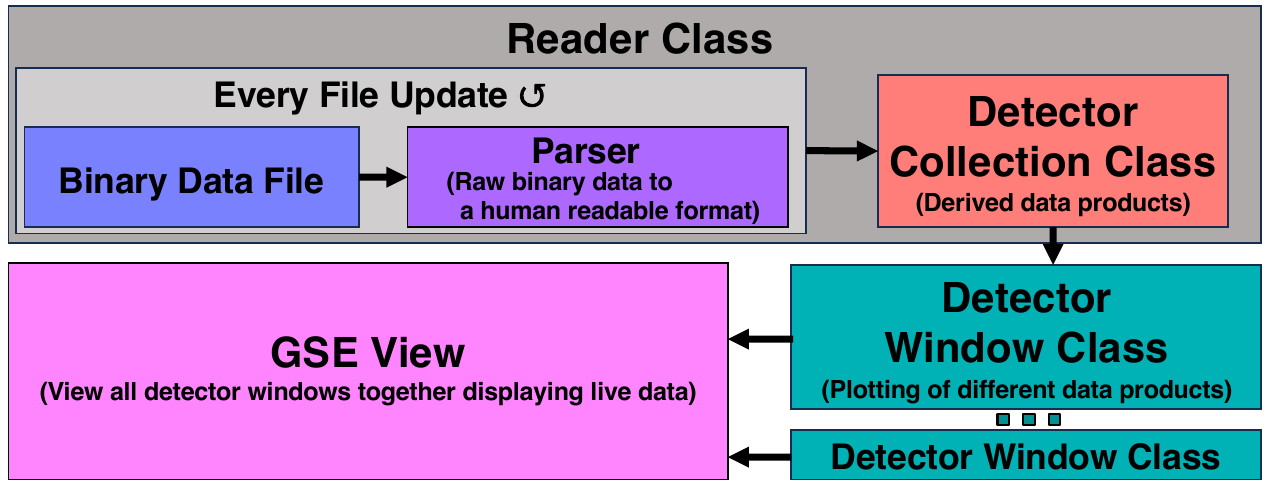} 
    \caption{The GSE data display pipeline. The top row indicates the reader class monitors for newly saved experiment data, is responsible for parsing that data, and manages how the new information is contained and accessed  using the detector-collection class. The information display for each detector is controlled via the detector-window class using the data from the reader class. All detector windows are then combined to form the full GSE display (see Fig~\ref{fig:gse_unit} left).}
    \label{fig:gse-vis-pipeline}
\end{figure}

As discussed in Section~\ref{sec:gse-data-logging}, telemetry data are saved to log files in a non-parsed binary format for each system. As shown in Fig~\ref{fig:gse-vis-pipeline}, a Reader Class for each log file is created to monitor and manage newly saved data. Once data is appended to a file, the Reader Class will pass that raw data to that system's parser which outputs the new data in a human-readable format. The parsed contents are then controlled by a detector-collection class that allows easy access to the new data and derived higher-level products (e.g., values with converted units or image products). 
A detector-window class is created for each individual component required for the display (see Fig~\ref{fig:gse_unit} left); for example, each detector, a display control box, and a whole-system log window. The detector-window class controls how the new data from the reader class is displayed. Once a detector-window is created for each component of the FOXSI-4 experiment, they are then combined into a single GUI element in order to display data being taken from the experiment in real-time. The data display performed well during flight, with the frame rate being limited to the data logging. The interactive button to restart image integration for each detector also worked as expected, this was crucial for fine-pointing during flight to ensure the flaring source was adequately positioned in the field of view of each telescope.

\section{Summary and Conclusion} 
We developed a modular, high-speed, and scalable data acquisition system for the FOXSI-4 and upcoming FOXSI-5 sounding rocket experiments, which are designed to perform direct focusing X-ray observations of solar flares with high temporal and spectral resolution.
To address the diverse requirements posed by multiple detector types and high photon flux conditions during flares, the system used a SpaceWire-based communication network with RMAP and distributed readout electronics (Section \ref{sec:daq_foxsi4}).

For FOXSI-4, we demonstrated successful integration and operation of three distinct detector systems: four CdTe-DSDs (Section \ref{sec:CdTe-DE_overview}), two CMOS sensors (Section \ref{sec:cmos_daq}), and a Quad-Timepix3 detector (Section \ref{sec:timepix_daq}).
Each system was paired with a dedicated readout module capable of robust control, data acquisition, and telemetry preparation.
To support integration with the SpaceWire-based communication network, we also developed two FPGA-based boards: SPMU-001 (Section \ref{sec:spmu001}) and SPMU-002 (Section \ref{sec:spmu002}). Both were designed as general-purpose platforms for physical measurement systems, with onboard SDRAM, Ethernet interface, and full support for SpaceWire and RMAP. These boards provided a unified and scalable interface for a range of detector subsystems, enabling efficient and modular integration without the need for instrument-specific hardware.
A central Formatter unit (Section \ref{sec:formatter}) coordinated real-time data and command flows, supported by a GSE (Section \ref{sec:gse}) for live system monitoring, uplink command control, and data logging.

The April 2024 FOXSI-4 flight, which successfully observed a GOES M1.6 solar flare, demonstrated the end-to-end robustness of the DAQ system: all six telescopes equipped with CdTe-DSD or CMOS sensors collected valuable science data, uplink commanding remained responsive throughout the flight, and QL data, including images, spectra, and housekeeping information, were successfully transmitted and displayed on the ground in real time. 
Although the Quad-Timepix3 detector did not return reliable data during FOXSI-4, post-flight analysis and bench tests indicate that the underlying cause was improper ASIC configuration triggered by the FPGA power-cycle sequence. Ground testing reproduced this behavior and showed that reliable operation required delaying power cycles and limiting configuration to a single load. For FOXSI-5, the system now includes updated startup checks, expanded telemetry to verify ASIC configuration, and a bleed resistor that discharges the FPGA within a few seconds.
%Although the Timepix3 detector did not record reliable in-flight data, its DAQ subsystem had been fully integrated and verified during preflight testing.
% Although the Timepix3 detector did not record reliable in-flight data, considered to be due to a power-related issue interfering with proper startup, its DAQ subsystem had been fully integrated and verified during preflight testing.
Building on these results, the same DAQ framework will be reused and further refined for the FOXSI-5 mission, scheduled for launch in 2025–2026.
The modular and standardized design, along with the development of the FPGA boards and open-source software libraries, provides a versatile template for future small satellite and suborbital missions that must integrate multiple detector technologies within operational and resource constraints.

% \clearpage
\appendix    % this command starts appendixes

\section{Design of Data Structure and Operation Mode of CdTe-DSD}\label{sec:design_data}
\begin{figure}[h]
    \begin{center}
    \includegraphics[width=1.0\hsize]{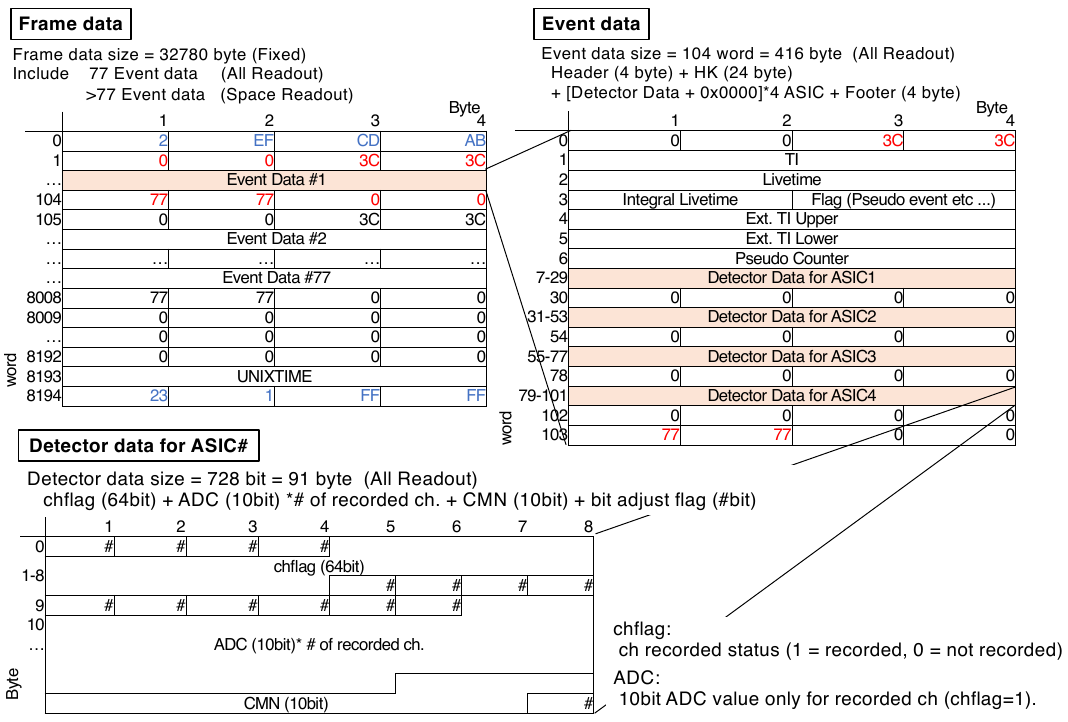}
    \end{center}
    \caption{Binary data structure of CdTe-DSD. Data is read/written in a ring buffer in ``Frame data" unit. In Frame data, more than 77 ``Event data" are included. In Event data, housekeeping information (FPGA clock counter TI, livetime, and pseudo event flag) and detector data (ch, ADC, and common-mode noise) for each ASIC channel are included. In Sparse Readout Mode, only channels whose ADC value exceeds the digital threshold $D_{th}$ are read out and managed by chflag.}
    \label{data_format}
\end{figure}

Fig\,\ref{data_format} shows a binary data structure of a CdTe-DSD. Writing to and reading data from the ring buffer is done in units of data called ``Frame data", with a fixed data size of 32 kB for easier handling.
Frame data contains several ``Event data", which include information on the channel, ADC, and common-mode noise recorded by each ASIC, as well as housekeeping information such as the FPGA clock counter (TI) and livetime. 
In the case of ``All Readout Mode" in ASIC (see Section \ref{sec:data_max}), the ADCs of all 256 channels are recorded, so the event data size is fixed (416 bytes), and 77 events are included per frame. In the case of ``Sparse Readout Mode" in ASIC, only the channels that exceed a certain ADC value are read out (the digital threshold parameter $D_{th}$ in ASIC), so the event data size becomes smaller, and the number of events included in one frame is therefore larger.

\begin{figure}[h]
  \begin{center}
  \includegraphics[width=1.0\hsize]{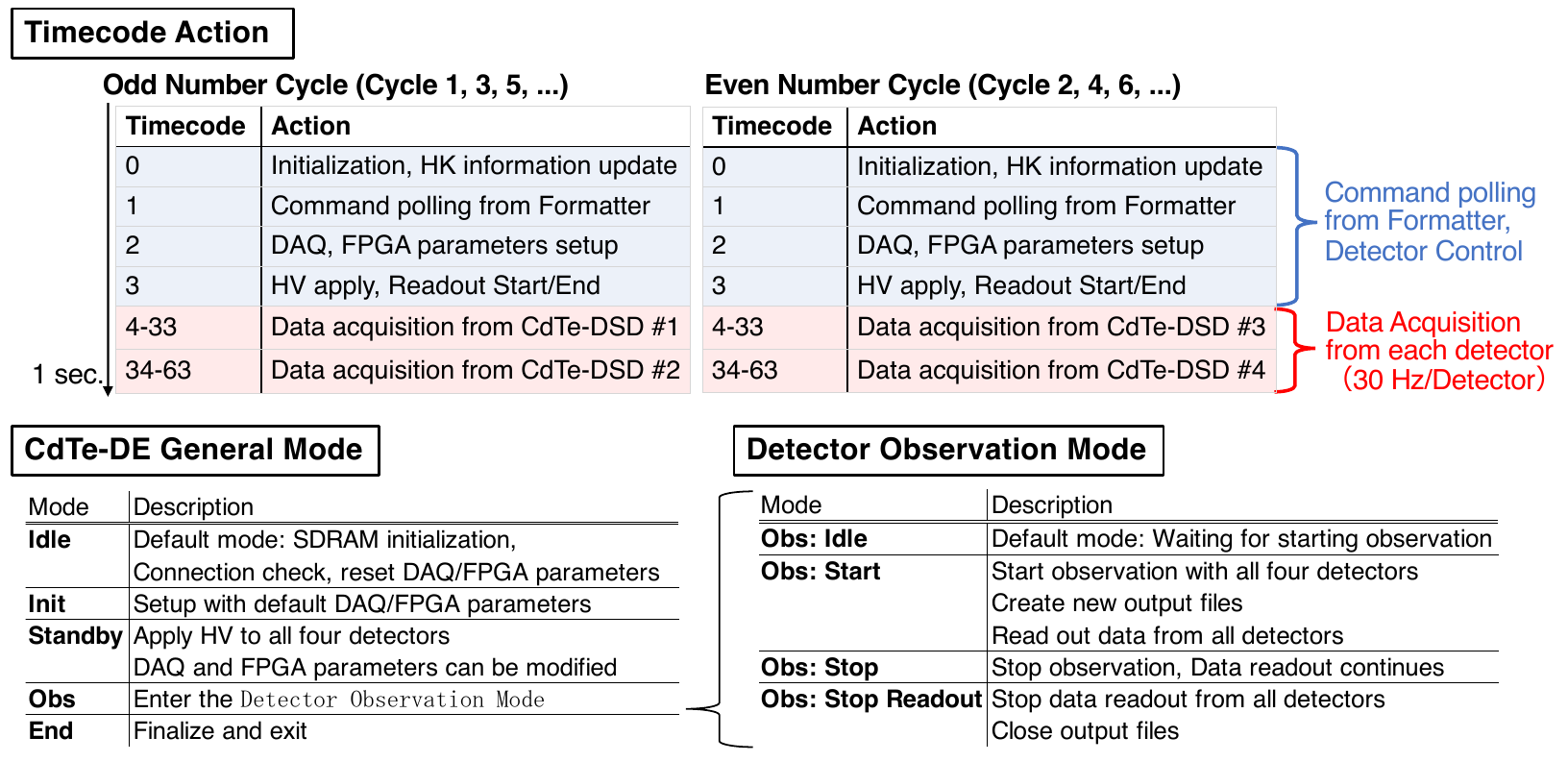}
  \end{center}
  \caption{(Upper): Timecode action for the CdTe-DE. (Lower): Operation modes of the CdTe-DE}
  \label{data_timecode}
\end{figure}

SpaceWire Timecode is a clock value broadcast over the SpaceWire network, consisting of increment values from 0 to 63, and with a period of 64 Hz. The CdTe-DE generates and monitors the Timecode value to schedule its control and data acquisition tasks.
As shown in Fig\,\ref{data_timecode} upper, the Timecode sequence is divided into even- and odd-numbered cycles, with each one-second interval consisting of 64 Timecode steps (1 step = 1/64 s = 15.6 ms). Of these, 4 Hz is allocated for receiving commands from the Formatter and updating the configuration of the CdTe-DSDs, while the remaining 60 Hz is dedicated to data acquisition.
Specifically, detectors \#1 and \#2 are read out during even-numbered cycles, while detectors \#3 and \#4 are read out during odd-numbered cycles.
This structured timing scheme enables efficient communication, minimizes latency, and ensures load-balanced data acquisition from all four detectors.
For simpler control, two layers of operation modes are defined: {\tt CdTe-DE General Mode}, which governs overall system behavior, and {\tt Detector Observation Mode}, which manages the details of data acquisition states of the detector.
Commands from the Formatter can modify these operation modes as well as key detector parameters such as the high voltage (HV) and thresholds for event triggers. These commands are executed by overwriting the contents of a 12-byte command buffer located in the SDRAM of CdTe-DE.

Fig\,\ref{data_timecode} lower summarizes the operation modes of CdTe-DE.
The {\tt CdTe-DE General Mode} comprises five states: {\tt Idle}, {\tt Init}, {\tt Standby}, {\tt Obs}, and {\tt End}. Upon power-on, the system uses systemd to automatically launch the CdTe-DE software, entering the {\tt Idle} state. In this mode, communication with the detectors is verified and the SDRAM and ASICs are initialized.
The system then transitions to {\tt Init}, where lab-calibrated default parameters are applied.
In {\tt Standby}, bias voltages are enabled, and ASIC and data acquisition parameters can be adjusted if needed (e.g., to mitigate excess noise).
During {\tt Obs}, observation-related settings are configured, and detailed control is delegated to the {\tt Detector Observation Mode}. The {\tt End} mode halts all detector operations and returns the system to {\tt Standby}.
The {\tt Detector Observation Mode}, active only while {\tt CdTe-DE General Mode} is set to {\tt Obs}, comprises four sub-states: {\tt Obs:Idle}, {\tt Obs:Start}, {\tt Obs:Stop}, and {\tt Obs:Stop Readout}.
The system begins in {\tt Obs:Idle} and transitions to {\tt Obs:Start} once all four detectors initiate data acquisition.
Detector data are written into the ring buffer of each detector electronics canister and simultaneously retrieved by the CdTe-DE during Timecodes 4–63, then saved as binary files on the Raspberry Pi.
In {\tt Obs:Stop}, detector acquisition ends, but the CdTe-DE continues retrieving residual data from the ring buffers. Finally, {\tt Obs:Stop Readout} terminates all readout operations and returns the system to {\tt Obs:Idle}.

\subsection*{Disclosures}
The authors declare that there are no financial interests, commercial affiliations, or other potential conflicts of interest that could have influenced the objectivity of this research or the writing of this paper

\subsection* {Code, Data, and Materials Availability} 
Data are available from the authors upon request.
Software is publicly available from the FOXSI GitHub repository: \url{https://github.com/foxsi}.

\subsection* {Acknowledgments}
The FOXSI-4 and FOXSI-5 sounding rocket missions are funded by NASA LCAS grant 80NSSC22K0577 and 80NSSC25M7045, respectively. We acknowledge the NSF for its support of space physics at UMN via grants AGS-1429512 and 1752268. This work was also supported by JSPS KAKENHI Grant Numbers JP18H03724, JP21H04486, JP21KK0052, JP22H00134, and JP22J12583.
This work was also supported by ISAS program for small-scale projects and the joint research program of the Institute for Space-Earth Environmental Research, Nagoya University.
The FOXSI team is grateful to the NASA and NSROC teams at WSMR and Wallops, and the staff of the Poker Flat Research Range and UAF Geophysical Institute, for supporting this mission.
Furthermore, the authors would like to acknowledge the contributions of each member of the FOXSI experiment team to the project, across the many institutions, areas of expertise, and career level of this broad and diverse team.
The authors also would like to acknowledge Shimafuji Electric Incorporated for the development of the FPGA boards SPMU-001 and SPMU-002.
S.N. is supported by FoPM (WISE Program) and JSR Fellowship, The University of Tokyo, Japan, and by the Overseas Research Fellowship by JSPS.
S.N. is also supported by a grant from the Hayakawa Satio Fund awarded by the Astronomical Society of Japan.
During the preparation of this manuscript, the authors used ChatGPT-4 to improve readability and language. After using this tool, the authors reviewed and edited the content as necessary and take full responsibility for the final version of the publication.

%%%%% References %%%%%
\begin{spacing}{0.95}
\makeatletter
\let\spie@thebibliography\thebibliography
\renewcommand{\thebibliography}[1]{%
  \spie@thebibliography{#1}%
  \setlength{\itemsep}{0.15\baselineskip}% ← ここでRef間隔を調整
  \setlength{\parskip}{0pt}% spieman.cls の負の parskip を打ち消す
  \setlength{\parsep}{0pt}%
}
\makeatother
\bibliography{main}   % bibliography data in main.bib

@article{hurford2003rhessi,
  title={{The RHESSI imaging concept}},
  author={Hurford, GJ c and Schmahl, EJ and Schwartz, RA and Conway, AJ and Aschwanden, MJ and Csillaghy, A and Dennis, BR and Johns-Krull, C and Krucker, S and Lin, RP and others},
  journal={The Reuven Ramaty High-Energy Solar Spectroscopic Imager (RHESSI) Mission Description and Early Results},
  pages={61--86},
  year={2003},
  publisher={Springer}
}

@book{lin2003reuven,
  title={{The Reuven Ramaty high-energy solar spectroscopic imager (RHESSI)}},
  author={Lin, Robert P and Dennis, Brian R and Hurford, Gordon J and Smith, DM and Zehnder, Alex and Harvey, PR and Curtis, David W and Pankow, Dave and Turin, Paul and Bester, M and others},
  year={2003},
  publisher={Springer}
}

@book{judge2024problem,
  title={The Problem of Coronal Heating: A Rosetta Stone for Electrodynamic Coupling in Cosmic Plasmas},
  author={Judge, Philip and Ionson, James A},
  year={2024},
  publisher={Springer}
}

@article{krucker2014first,
  title     = {{First images from the focusing optics X-ray solar imager}},
  author    = {Krucker, S{\"a}m and Christe, Steven and Glesener, Lindsay and Ishikawa, Shinnosuke and Ramsey, Brian and Takahashi, Tadayuki and Watanabe, Shin and Saito, Shinya and Gubarev, Mikhail and Kilaru, Kiranmayee and others},
  journal   = {The Astrophysical Journal Letters},
  volume    = {793},
  number    = {2},
  pages     = {L32},
  year      = {2014},
  publisher = {IOP Publishing}
}

@article{ishikawa2017detection,
  title     = {{Detection of nanoflare-heated plasma in the solar corona by the FOXSI-2 sounding rocket}},
  author    = {Ishikawa, Shinnosuke and Glesener, Lindsay and Krucker, S{\"a}m and Christe, Steven and Buitrago-Casas, Juan Camilo and Narukage, Noriyuki and Vievering, Juliana},
  journal   = {Nat. Astron.},
  volume    = {1},
  number    = {11},
  pages     = {771--774},
  year      = {2017},
  publisher = {Nature Publishing Group UK London}
}

@article{athiray2020foxsi,
  title     = {{FOXSI-2 solar microflares. I. Multi-instrument differential emission measure analysis and thermal energies}},
  author    = {Athiray, PS and Vievering, Juliana and Glesener, Lindsay and Ishikawa, Shinnosuke and Narukage, Noriyuki and Buitrago-Casas, Juan Camilo and Musset, Sophie and Inglis, Andrew and Christe, Steven and Krucker, S{\"a}m and others},
  journal   = {The Astrophysical Journal},
  volume    = {891},
  number    = {1},
  pages     = {78},
  year      = {2020},
  publisher = {IOP Publishing}
}

@article{vievering2021foxsi,
  title     = {{FOXSI-2 solar microflares. II. Hard X-ray imaging spectroscopy and flare energetics}},
  author    = {Vievering, Juliana T and Glesener, Lindsay and Athiray, PS and Buitrago-Casas, Juan Camilo and Musset, Sophie and Ryan, Daniel F and Ishikawa, Shinnosuke and Duncan, Jessie and Christe, Steven and Krucker, S{\"a}m},
  journal   = {The Astrophysical Journal},
  volume    = {913},
  number    = {1},
  pages     = {15},
  year      = {2021},
  publisher = {IOP Publishing}
}

@article{buitrago2022faintest,
  title     = {{The faintest solar coronal hard X-rays observed with FOXSI}},
  author    = {Buitrago-Casas, Juan Camilo and Glesener, Lindsay and Christe, Steven and Krucker, S{\"a}m and Vievering, Juliana and Athiray, PS and Musset, Sophie and Davis, Lance and Courtade, Sasha and Dalton, Gregory and others},
  journal   = {Astronomy \& Astrophysics},
  volume    = {665},
  pages     = {A103},
  year      = {2022},
  publisher = {EDP Sciences}
}

@techreport{burth2023nasa,
  title  = {{NASA Sounding Rockets User Handbook}},
  author = {Burth, Robert H and Cathell, Philip G and Edwards, David B and Ghalib, Ahmed H and Gsell, John C and Hales, Heath C and Haugh, Herbert C and Tibbetts, Brian R},
  year   = {2023}
}

@techreport{ecssSpaceWire,
  title = {{SpaceWire—links, nodes, routers, and networks}},
  author = {ECSS-E-ST-50-12C},
  year = {2019},
  type = {Standard},
  institution = {ESA-ESTEC Requirements and Standards Division}
}

@techreport{ecssRMAP,
  title = {{SpaceWire—Remote Memory Access Protocol}},
  author = {ECSS-E-ST-50-52C},
  year = {2010},
  type = {Standard},
  institution = {ESA-ESTEC Requirements and Standards Division}
}

@article{nagasawa2023wide,
  title={{Wide-gap CdTe strip detectors for high-resolution imaging in hard X-rays}},
  author={Nagasawa, Shunsaku and Minami, Takahiro and Watanabe, Shin and Takahashi, Tadayuki},
  journal={Nuclear Instruments and Methods in Physics Research Section A: Accelerators, Spectrometers, Detectors and Associated Equipment},
  volume={1050},
  pages={168175},
  year={2023},
  publisher={Elsevier}
}

@article{ISHIKAWA2018191,
title = {{High-speed X-ray imaging spectroscopy system with Zynq SoC for solar observations}},
journal = {Nuclear Instruments and Methods in Physics Research Section A: Accelerators, Spectrometers, Detectors and Associated Equipment},
volume = {912},
pages = {191-194},
year = {2018},
note = {New Developments In Photodetection 2017},
issn = {0168-9002},
doi = {https://doi.org/10.1016/j.nima.2017.11.033},
author = {Shinnosuke Ishikawa and Tadayuki Takahashi and Shin Watanabe and Noriyuki Narukage and Satoshi Miyazaki and Tadashi Orita and Shin’ichiro Takeda and Masaharu Nomachi and Iwao Fujishiro and Fumio Hodoshima}
}

@inproceedings{shimizu2024evaluation,
  title={{Evaluation of a CMOS sensor for solar flare soft x-ray imaging spectroscopy onboard the sounding rocket experiment FOXSI-4}},
  author={Shimizu, Riko and Narukage, Noriyuki and Sakao, Taro and Sato, Yoshiaki and Kashima, Sota and Takahashi, Tadayuki and Nagasawa, Shunsaku and Minami, Takahiro and Iwata, Toshiya and Glesener, Lindsay and others},
  booktitle={X-Ray, Optical, and Infrared Detectors for Astronomy XI},
  volume={13103},
  pages={63--78},
  year={2024},
  organization={SPIE}
}

@inproceedings{savage2022first,
  title     = {{The First Solar Flare Sounding Rocket Campaign}},
  author    = {Savage, Sabrina and Winebarger, Amy and Glesener, Lindsay and Reves, Katharine and Kobayashi, Ken and Golub, Leon and Chamberlain, Phil},
  booktitle = {NASA Sounding Rocket Symposium 2022},
  year      = {2022}
}

@inproceedings{krucker2013focusing,
  title={{The focusing optics x-ray solar imager (FOXSI): instrument and first flight}},
  author={Krucker, S{\"a}m and Christe, Steven and Glesener, Lindsay and Ishikawa, Shinnosuke and Ramsey, Brian and Gubarev, Mikhail and Saito, Shinya and Takahashi, Tadayuki and Watanabe, Shin and Tajima, Hiroyasu and others},
  booktitle={Solar Physics and Space Weather Instrumentation V},
  volume={8862},
  pages={261--272},
  year={2013},
  organization={SPIE}
}

@article{buitrago2022quad,
  title={{The Quad-Timepix/Timepix3 Detector: A novel solution for coming solar high energy space experiments}},
  author={Buitrago-Casas, Juan and Perez-Piel, Savannah and Tremsin, Anton and Oliveros, Juan Carlos Martinez},
  journal={Third Triennial Earth-Sun Summit (TESS)},
  volume={54},
  number={7},
  year={2022}
}

@inproceedings{yuasa2008portable,
  title={{A Portable SpaceWire/RMAP Class Library for Scientific Detector Read Out Systems}},
  author={Yuasa, Takayuki and Kokuyama, Wataru and Makishima, Kazuo and Nakazawa, Kazuhiro and Nomachi, Masaharu and Kokubun, Motohide and Odaka, Hirokazu and Takashima, Takeshi and Takahashi, Tadayuki},
  booktitle={International SpaceWire Conference},
  pages={4--6},
  year={2008}
}

@article{kokubun2007orbit,
    author = {Kokubun, Motohide and Makishima, Kazuo and Takahashi, Tadayuki and Murakami, Toshio and Tashiro, Makoto and Fukazawa, Yasushi and Kamae, Tuneyoshi and Madejski, Greg M. and Nakazawa, Kazuhiro and Yamaoka, Kazutaka and Terada, Yukikatsu and Yonetoku, Daisuke and Watanabe, Shin and Tamagawa, Toru and Mizuno, Tsunefumi and Kubota, Aya and Isobe, Naoki and Takahashi, Isao and Sato, Goro and Takahashi, Hiromitsu and Hong, Soojing and Kawaharada, Madoka and Kawano, Naomi and Mitani, Takefumi and Murashima, Mio and Suzuki, Masaya and Abe, Keiichi and Miyawaki, Ryouhei and Ohno, Masanori and Tanaka, Takaaki and Yanagida, Takayuki and Itoh, Takeshi and Ohnuki, Kousuke and Tamura, Ken-ichi and Endo, Yasuhiko and Hirakuri, Shinya and Hiruta, Tatsuro and Kitaguchi, Takao and Kishishita, Tetsuichi and Sugita, Satoshi and Takahashi, Takuya and Takeda, Shin’ichiro and Enoto, Teruaki and Hirasawa, Ayumi and Katsuta, Jun’ichiro and Matsumura, Satoshi and Onda, Kaori and Sato, Mitsuhiro and Ushio, Masayoshi and Ishikawa, Shin-nosuke and Murase, Koichi and Odaka, Hirokazu and Suzuki, Masanobu and Yaji, Yuichi and Yamada, Shinya and Yamasaki, Tomonori and Yuasa, Takayuki and the HXD team},
    title = {{In-Orbit Performance of the Hard X-Ray Detector on Board Suzaku}},
    journal = {Publications of the Astronomical Society of Japan},
    volume = {59},
    number = {sp1},
    pages = {S53-S76},
    year = {2007},
    month = {01},
    issn = {0004-6264},
    doi = {10.1093/pasj/59.sp1.S53},
    url = {https://doi.org/10.1093/pasj/59.sp1.S53},
}

@article{nagasawa2025imaging,
  title={{Imaging and spectral performance of a wide-gap CdTe double-sided strip detector}},
  author={Nagasawa, Shunsaku and Minami, Takahiro and Watanabe, Shin and Takahashi, Tadayuki},
  journal={Nuclear Instruments and Methods in Physics Research Section A: Accelerators, Spectrometers, Detectors and Associated Equipment},
  volume={1075},
  pages={170362},
  year={2025},
  publisher={Elsevier}
}

@article{watanabe2014si,
  title     = {{The Si/CdTe semiconductor Compton camera of the ASTRO-H soft gamma-ray detector (SGD)}},
  author    = {Watanabe, Shin and Tajima, Hiroyasu and Fukazawa, Yasushi and Ichinohe, Yuto and Enoto, Teruaki and Fukuyama, Taro and Furui, Shunya and Genba, Kei and Hagino, Kouichi and Harayama, Atsushi and others},
  journal   = {NIMA},
  volume    = {765},
  pages     = {192--201},
  year      = {2014},
 doi = {https://doi.org/10.1016/j.nima.2014.05.127},
  publisher = {Elsevier}
}

@inproceedings{ozaki2010spacewire,
  title     = {{SpaceWire driven architecture for the ASTRO-H satellite}},
  author    = {Ozaki, Masanobu and Takahashi, Tadayuki and Kokubun, Motohide and Takashima, Takeshi and Odaka, Hirokazu and Nomachi, Masaharu and Yuasa, Takayuki and Fujishiro, Iwao and Tohma, Takayuki and Hihara, Hiroki and others},
  booktitle = {Proc. of the 3rd Int. SpaceWire Conf},
  volume    = {445},
  year      = {2010}
}

@inproceedings{hihara2020onboard,
  title     = {{Onboard signal processing system of HAYABUSA2}},
  author    = {Hihara, Hiroki and Sano, Junpei and Takada, Jun and Masuda, Tetsuya and Okada, Tatsuaki and Ogawa, Naoko and Ootake, Hisashi and Tsuda, Yuichi},
  booktitle = {Proc. SPIE},
  volume    = {11502},
  pages     = {124--139},
  year      = {2020}
}

@inproceedings{rakow2003reliable,
  title        = {{Reliable transport over SpaceWire for James Webb space telescope}},
  author       = {Rakow, G and Schmirr, R and Dailey, C and Shakoorzadeh, K},
  booktitle    = {IEEE Aerosp. Conf. Proc.},
  volume       = {5},
  pages        = {2289--52302},
  year         = {2003},
  organization = {IEEE}
}

@inproceedings{buitrago2021foxsi,
author = {Juan Camilo Buitrago-Casas and Juliana Vievering and Sophie Musset and Lindsay Glesener and P.S. Athiray and Wayne Baumgartner and Stephen Bongiorno and Patrick Champey and Steven Christe and Sasha Courtade and Gregory Dalton and Jessie Duncan and Kelsey Gilchrist and Shin-nosuke Ishikawa and Christine Jhabvala and Hunter Kanniainen and Sam Krucker and Kyle Gregory and Juan Carlos Martinez Oliveros and Jeff McCracken and Ikuyuki Mitsuishi and Noriyuki Narukage and Athanasios Pantazides and Eliad Peretz and Savannah Perez-Piel and Aruna Ramanayaka and Brian Ramsey and Danny Ryan and Sabrina Savage and Tadayuki Takahashi and Shin Watanabe and Amy Winebarger and Yixian Zhang},
title = {{FOXSI-4: the high resolution focusing X-ray rocket payload to observe a solar flare}},
volume = {11821},
booktitle = {UV, X-Ray, and Gamma-Ray Space Instrumentation for Astronomy XXII},
editor = {Oswald H. Siegmund},
organization = {International Society for Optics and Photonics},
publisher = {SPIE},
pages = {118210L},
year = {2021},
doi = {10.1117/12.2594701},
URL = {https://doi.org/10.1117/12.2594701}
}

@article{Vallerga2011,
doi = {10.1088/1748-0221/6/01/C01049},
url = {https://dx.doi.org/10.1088/1748-0221/6/01/C01049},
year = {2011},
month = {jan},
volume = {6},
number = {01},
pages = {C01049},
author = {J Vallerga and R Raffanti and A Tremsin and J McPhate and O Siegmund},
title = {{MCP detector read out with a bare quad Timepix at kilohertz framerates}}
}

@article{McCarter23,
  title = {{Antiferromagnetic Real-space Configuration Probed by Dichroism in Scattered X-ray Beams with Orbital Angular Momentum}},
  author = {McCarter, Margaret R. and Saleheen, Ahmad I. U. and Singh, Arnab and Tumbleson, Ryan and Woods, Justin S. and Tremsin, Anton S. and Scholl, Andreas and De Long, Lance E. and Hastings, J. Todd and Morley, Sophie A. and Roy, Sujoy},
  journal = {Phys. Rev. B},
  volume = {107},
  issue = {6},
  pages = {L060407},
  numpages = {7},
  year = {2023},
  month = {Feb},
  publisher = {American Physical Society},
  doi = {10.1103/PhysRevB.107.L060407},
  url = {https://link.aps.org/doi/10.1103/PhysRevB.107.L060407}
}

@article{Tremsin21,
author = "Tremsin, Anton S. and Vallerga, John V. and Siegmund, Oswald H. W. and Woods, Justin and De Long, Lance E. and Hastings, Jeffrey T. and Koch, Roland J. and Morley, Sophie A. and Chuang, Yi-De and Roy, Sujoy",
title = {{Photon-counting MCP/Timepix detectors for soft X-ray imaging and spectroscopic applications}},
journal = "Journal of Synchrotron Radiation",
year = "2021",
volume = "28",
number = "4",
pages = "1069--1080",
month = "Jul",
doi = {10.1107/S1600577521003908},
url = {https://doi.org/10.1107/S1600577521003908}
}

@article{Pitters19,
doi = {10.1088/1748-0221/14/05/P05022},
url = {https://dx.doi.org/10.1088/1748-0221/14/05/P05022},
year = {2019},
month = {may},
volume = {14},
number = {05},
pages = {P05022},
author = {Pitters, F. and Tehrani, N. Alipour and Dannheim, D. and Fiergolski, A. and Hynds, D. and Klempt, W. and Llopart, X. and Munker, M. and Nürnberg, A. and Spannagel, S. and Williams, M.},
title = {{Time resolution studies of Timepix3 assemblies with thin silicon pixel sensors}},
journal = {Journal of Instrumentation}
}

@article{Burian19,
doi = {10.1088/1748-0221/14/01/C01001},
url = {https://dx.doi.org/10.1088/1748-0221/14/01/C01001},
year = {2019},
month = {jan},
volume = {14},
number = {01},
pages = {C01001},
author = {Burian, P. and Broulím, P. and Bergmann, B.},
title = {{Study of Power Consumption of Timepix3 Detector}},
journal = {Journal of Instrumentation}
}

@article{Gaspari14,
doi = {10.1088/1748-0221/9/01/C01037},
url = {https://dx.doi.org/10.1088/1748-0221/9/01/C01037},
year = {2014},
month = {jan},
publisher = {},
volume = {9},
number = {01},
pages = {C01037},
author = {M De Gaspari and J Alozy and R Ballabriga and M Campbell and E Fröjdh and J Idarraga and S Kulis and X Llopart and T Poikela and P Valerio and W Wong},
title = {{Design of the analog front-end for the Timepix3 and Smallpix hybrid pixel detectors in 130 nm CMOS technology}},
journal = {Journal of Instrumentation}
}

@manual{tshark,
  author       = {{Wireshark Foundation}},
  title        = {{TShark: The Wireshark Network Analyzer – Command-Line Version}},
  organization = {Wireshark Foundation},
  note         = {Accessed May 2025},
  url          = {https://www.wireshark.org/docs/man-pages/tshark.html},
  year         = {2025}
}

@book{piana2022hard,
  title={{Hard X-ray imaging of solar flares}},
  author={Piana, Michele and Emslie, A Gordon and Massone, Anna Maria and Dennis, Brian R},
  volume={164},
  year={2022},
  publisher={Springer}
}

@article{NARUKAGE2020162974,
title = {{High-speed back-illuminated CMOS sensor for photon-counting-type imaging-spectroscopy in the soft X-ray range}},
journal = {Nuclear Instruments and Methods in Physics Research Section A: Accelerators, Spectrometers, Detectors and Associated Equipment},
volume = {950},
pages = {162974},
year = {2020},
issn = {0168-9002},
doi = {https://doi.org/10.1016/j.nima.2019.162974},
url = {https://www.sciencedirect.com/science/article/pii/S0168900219313609},
author = {Noriyuki Narukage and Shinnosuke Ishikawa and Taro Sakao and Xinyang Wang}
}

@misc{SpaceWireRMAPLibrary,
  author       = {Yuasa, Takayuki},
  title        = {{SpaceWire RMAP Library version 2}},
  howpublished = {\url{https://github.com/yuasatakayuki/SpaceWireRMAPLibrary}}
}

@misc{CxxUtility,
  author       = {Yuasa, Takayuki},
  title        = {{CxxUtilities}},
  howpublished = {\url{https://github.com/yuasatakayuki/CxxUtilities}}
}

@misc{Formatter_soft,
  author       = {Pantazides, Athanasios and {FOXSI-4 and FOXSI-5 team}},
  title        = {{Formatter software, Version 1.2.0}},
  howpublished = {\url{https://github.com/foxsi/foxsi-4matter/releases/v1.2.0}}
}

@misc{Boost_Asio,
  author       = {Kohlhoff, Christopher M.},
  title        = {boost C++ libraries: Boost.Asio},
  howpublished = {\url{https://www.boost.org/doc/libs/latest/doc/html/boost_asio.html}}
}

@misc{GSE_soft,
  author       = {Cooper, Kristopher and {FOXSI-4 and FOXSI-5 team}},
  title        = {{GSE software, Version 3.0.0}},
  howpublished = {\url{https://github.com/foxsi/GSE-FOXSI-4/releases/tag/v3.0.0}}
}

@article{goto2001construction,
  title={{Construction and commissioning of a 215-m-long beamline at SPring-8}},
  author={Goto, S and Takeshita, K and Suzuki, Y and Ohashi, H and Asano, Y and Kimura, H and Matsushita, T and Yagi, N and Isshiki, M and Yamazaki, H and others},
  journal={Nuclear instruments and methods in physics research section A: accelerators, spectrometers, detectors and associated equipment},
  volume={467},
  pages={682--685},
  year={2001},
  publisher={Elsevier}
}
\bibliographystyle{spiejour}   % makes bibtex use spiejour.bst
\end{spacing}

%%%%% Biographies of authors %%%%%
% \noindent Biographies of the authors are not available.

% \end{spacing}
\end{document}